\documentclass[twocolumn,trackchanges]{aastex701}

\usepackage{amsmath}
\usepackage{threeparttable}
\usepackage{caption}

\begin{document}

\title{Modeling extraordinary 
relativistic jets on the scale of the cosmic web}

\author[orcid=0000-0001-9133-1005,gname=Gourab,sname='Giri']{Gourab Giri}
\affiliation{Istituto Nazionale di Astrofisica (INAF) - Istituto di Radioastronomia (IRA), Via Gobetti 101, I-40129 Bologna, Italy}
\email[show]{g.giri@ira.inaf.it}  

\author{Marco Tavani}
\affiliation{Istituto Nazionale di Astrofisica (INAF) -- Istituto di Astrofisica e Planetologia Spaziali (IAPS), via del Fosso del Cavaliere 100, Roma, 00133, Italy}
\email{marco.tavani@inaf.it}

\author{Dario Borgogno}
\affiliation{Istituto Nazionale di Astrofisica (INAF) -- Istituto di Astrofisica e Planetologia Spaziali (IAPS), via del Fosso del Cavaliere 100, Roma, 00133, Italy}
\email{dario.borgogno@inaf.it}

\author[orcid=0000-0002-9325-1567]{Andrea Botteon}
\affiliation{Istituto Nazionale di Astrofisica (INAF) - Istituto di Radioastronomia (IRA), Via Gobetti 101, I-40129 Bologna, Italy}
\email{andrea.botteon@inaf.it} 

\author{Rossella Cassano}
\affiliation{Istituto Nazionale di Astrofisica (INAF) - Istituto di Radioastronomia (IRA), Via Gobetti 101, I-40129 Bologna, Italy}
\email{rossella.cassano@inaf.it} 

\author{Valerio Vittorini}
\affiliation{Istituto Nazionale di Astrofisica (INAF) -- Istituto di Astrofisica e Planetologia Spaziali (IAPS), via del Fosso del Cavaliere 100, Roma, 00133, Italy}
\email{valerio.vittorini@inaf.it}

\author[orcid=0000-0003-1063-3541,gname=Claudio, sname='Gheller']{Claudio Gheller} 
\affiliation{Istituto Nazionale di Astrofisica (INAF) - Istituto di Radioastronomia (IRA), Via Gobetti 101, I-40129 Bologna, Italy}
\email{claudio.gheller@inaf.it}

\author[orcid=0000-0003-4195-8613]{Gianfranco Brunetti}
\affiliation{Istituto Nazionale di Astrofisica (INAF) - Istituto di Radioastronomia (IRA), Via Gobetti 101, I-40129 Bologna, Italy}
\email{gianfranco.brunetti@inaf.it}

\author[orcid=0000-0002-1029-3746]{Paul J. Wiita}
\affiliation{Department of Physics, The College of New Jersey, 2000 Pennington Road, Ewing, NJ, 08628-0718, USA}
\email{wiitap@tcnj.edu}

\author{Tiziana Venturi}
\affiliation{Istituto Nazionale di Astrofisica (INAF) - Istituto di Radioastronomia (IRA), Via Gobetti 101, I-40129 Bologna, Italy}
\email{tiziana.venturi@inaf.it}

\author{Bernie Fanaroff}
\affiliation{South African Radio Astronomy Observatory, 2 Fir Street, Black River Park, Observatory, Cape Town 7925, South Africa}
\email{bfanaroff@sarao.ac.za}

\author[orcid=0000-0002-8352-6635,sname=Mignone,gname=Andrea]{Andrea Mignone}\affiliation{Dipartimento di Fisica, Università degli Studi di Torino, Via Pietro Giuria 1, Torino, I-10125, Italy }\email{andrea.mignone@unito.it}

\author[orcid=0000-0002-4464-8023]{D J Saikia}\affiliation{Fakultat f\"ur Physik, Universit\"at Bielefeld, Postfach 100131, D-33501 Bielefeld, Germany}\affiliation{Assam Don Bosco University, Guwahati, Assam, 781017, India}\email{dhrubasaikia.tifr.ccsu@gmail.com}

\author{Somak Raychaudhury}\affiliation{Department of Physics, Ashoka University, Sonipat, Haryana, 131029, India}\affiliation{Inter-University Centre for Astronomy and Astrophysics, Ganeshkhind, Pune, 411007, India}\affiliation{School of Physics and Astronomy, University of Birmingham, Birmingham, B15 2TT, UK}\email{somak.raychaudhury@ashoka.edu.in}

\begin{abstract}

Recent observations of supermassive black hole jets extending up to the typical sizes of cosmic web filaments challenge our current understanding. In particular, the overall jet size of J152932.16+601534.4 (``Porphyrion'') reaches $\sim 7$ Mpc at redshift $0.896$. Magneto-hydrodynamic (MHD) models of relativistic jets have never been successfully applied to such extraordinary distances, requiring demanding jet-environment and flow stability conditions. By using state-of-the-art numerical simulations of 3D relativistic-MHD jets, we show that Porphyrion's observational properties can be successfully modelled in a realistic scenario, involving a jet kinetic power of $3.5 \times 10^{46}\, \rm erg \, s^{-1}$ and an external medium of a low-density galaxy group. We find that the pair of jets expands to $\sim 9$~Mpc in about $20$~Myr, implying that the jet grows with relativistic speeds for most of its evolution.
Comparing this case with other jets of similar power propagating in substantially denser cosmic structures, we find much slower outflows that can only reach a few Mpc, consistent with the bulk of the observed population of giant radio galaxies. 
Our results make a testable prediction, i.e. future detections of even larger-scale jets in upcoming surveys and database searches will be invaluable in studying the demographics and feedback
of the poorly known population of powerful jets, particularly in under-dense regions of the cosmic web.

\end{abstract}



\section{Introduction} 

Which physical processes govern the maximum size a relativistic extragalactic jet can attain remains an open question.
Observations in the 1970s provided the first evidence for extremely long jets,
most notably through the discovery of the radio galaxy 3C~236 \citep{Willis1974}, whose projected linear extent reaches 4.1~Mpc (for $H_0 = 70\ \mathrm{km\,s^{-1}\,Mpc^{-1}}$). Subsequent discoveries have steadily pushed this upper bound outward, and such sources are often called
``extreme'' giant radio galaxies (GRGs), with ordinary GRGs usually defined as spanning at least 0.7 Mpc \citep{Dabhade2023}. 
In 2008, J1420--0545 was reported to span 4.69~Mpc \citep{Machalski2008}, and more recently, in 2022, the GRG ``Alcyoneus'' (J081421.68+522410.0) was found to extend to $\sim 5.04$~Mpc \citep{Oei2022_5Mpc}.

The apparent clustering of the largest known radio sources around a characteristic scale of $\sim 5$~Mpc raises a fundamental question: why do the most powerful extragalactic sources appear to stall at this scale? Does this reflect an intrinsic limit to jet propagation imposed by dynamical instabilities, radiative losses, or interactions with the surrounding medium, or is it primarily governed by the structural extent and geometry of the large-scale cosmic environment through which these jets propagate? Alternatively, does it 
simply arise from an observational selection effect, such as incomplete sampling of low-surface-brightness extended structures? 

This apparent ceiling was decisively challenged in 2024 with the discovery of ``Porphyrion'' (J152932.16+601534.4), whose reported linear extent of $\sim 7$~Mpc exceeds previous records by a substantial margin \citep{Oei2024_7Mpc}. Further support for the existence of such large sources comes from giant radio quasars (GRQs). The largest known GRQ has a projected size of 4.45~Mpc \citep{Coziol2017}; however, within the framework of the quasar--radio galaxy unification scheme, quasars are expected to be observed at smaller angles to the line of sight than RGs. Adopting an inclination angle of $\sim 45^{\circ}$---often considered the dividing angle between RGs and quasars \citep{Barthel1989}---this projected size corresponds to an intrinsic extent of $\gtrsim 6.3$~Mpc. Moreover, recent 
radio interferometric surveys have identified a growing population of sources whose extent 
exceeds the previously assumed $\sim 5$~Mpc upper limit.  
Six GRGs with linear sizes surpassing 5~Mpc and reaching up to $\sim 6.6$~Mpc were recently reported \citep{Andernach2025}; \citep[also see,][]{Hardcastle2023}. 
These discoveries have revived the longstanding question: is there a fundamental physical limit to how far an extragalactic jet can propagate?

Studying these ``extreme" GRGs is also crucial, as such systems should embody all the 
conditions that enable sustained, large-scale jet growth, making them ideal laboratories for testing the limits of  jet propagation physics. A key question is 
whether such systems require prolonged, uninterrupted jet activity, or instead are they the result of multiple episodes of restarted jets \citep[e.g.,][]{Jamrozy2017}, with later outflows 
propagating more rapidly through an 
ambient medium made less dense by previous activity \citep{Ratnaparkhi2026,Dabhade2025}. Furthermore, the apparent ability of these jets to survive disruptive magneto-hydrodynamic (MHD) instabilities, combined with contradictory observational evidence on the influence of dense versus underdense cosmic environments, underscores significant gaps in our current understanding of the physical origin and evolution of these sources \citep{Tang2020,Dabhade2023,Sankhyayan2024,Andernach2025}. 

Beyond their dynamics, spatially extended jets represent a key frontier in feedback studies, functioning as conduits that transport energy across diverse scales to counter radiative cooling in cosmic media \citep{Husko2026}. However, it is unclear whether collimated, high-velocity jets deposit substantial energy directly into their immediate host environment, or if their strong momentum carries the bulk of their energetic output beyond the dense core and into the lower-density outskirts of the halo \citep{Hardcastle2013}.

In this context, the rapidly growing body of observations of GRGs \citep{Mostert2024} calls for a dedicated program of numerical simulations \citep[e.g.,][]{Giri2026,Giri2026_RGs,Giri2026_5Mpc} aimed at identifying the physical conditions, both intrinsic to the jet and imposed by the surrounding environment, that permit sustained jet propagation on multi-megaparsec scales. These systems represent the largest known structures produced by individual galaxies, occupying a regime of jet evolution that remains largely unexplored by current high-resolution MHD simulations.
Here we discuss 
a set of computationally demanding simulations designed to probe the growth conditions in which such ``extreme'' jet propagation can be realized. 

This article is organized as follows: Section~\ref{Sec:Numerical setup} describes the numerical setups for realistic jet propagation and modeling their radiative appearances. Section~\ref{Sec:Results} presents our main findings on jet propagation dynamics across varying environmental conditions. Section~\ref{Sec:Discussion} compares these results with observational data and discusses potential explanations for the rarity of extreme GRGs. Finally, Section~\ref{Sec:Summary} summarizes our key conclusions. 

\section{Numerical setup}\label{Sec:Numerical setup}

We performed our simulations using the \texttt{gPLUTO} code \citep{Rossazza2025}, the GPU-accelerated successor of the widely used \texttt{PLUTO} legacy code \citep{Mignone2007}, following its relativistic magneto-hydrodynamical formulation that was extensively validated in studies of astrophysical jets. The use of \texttt{gPLUTO} was essential to achieve the computational performance required for this work, and enabled simulations at a scale (Table~\ref{Tab:Simulation_setup}) that, to our knowledge, represents the most demanding zoom-in study of jet propagation to date \citep[as opposed to large-scale cosmological simulations with AGN feedback;][]{Husko2026}. The architecture of \texttt{gPLUTO} is described in detail in \citet{Rossazza2025}, while its performance has been benchmarked in \citet{Giri2026}, demonstrating speed-ups of approximately an order of magnitude on GPU-based computing systems, particularly for jet propagation problems. Following these papers, our simulations were performed in three-dimensional Cartesian grids (outflow boundaries on all sides, except for a reflective boundary at the jet-injection side) and adopted the Taub–Matthews equation of state \citep{Mignone2007_EoS}. To maintain the solenoidal condition of the magnetic field, we employed the divergence-cleaning method \citep{Dedner2002}. The simulations were carried out on the CINECA's Pitagora high-performance computing (HPC) facility \citep[e.g.,][]{Turisini2023}, one of the leading HPC infrastructures globally. We utilized 96 GPU nodes (each equipped with four \texttt{NVIDIA H100 SXM 94GB HBM2e} GPUs), with a typical simulation requiring $\sim 24$ hours of wall-clock time.

\begin{figure*}[h]
\centering
\includegraphics[width=\textwidth]{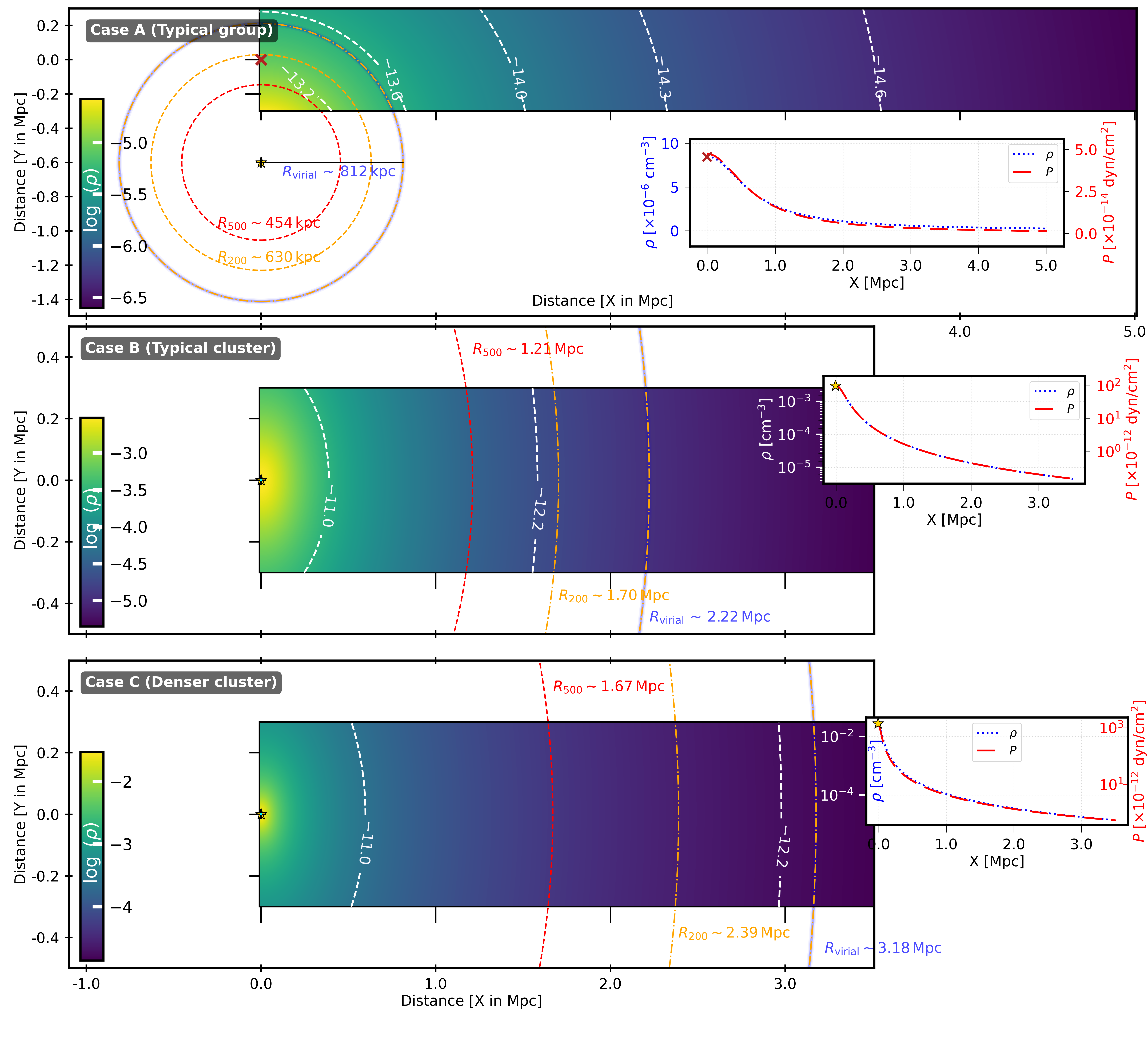}
\caption{2D slices ($x-y$; $z=0$) of 3D density ($\rho$) domains used for the three simulation cases. \textit{Top panel (``Case A''):} A powerful jet, injected horizontally in the $+x$ direction, from $x=y=0$ (location marked with `$\boldsymbol{\times}$') in the outskirts of a galaxy group (600 kpc above the group center) and propagates out of the group into a low-density cosmic void. \textit{Middle panel (``Case B''):} The same jet is injected at the center of a typical galaxy cluster (location marked with `$\bigstar$').
\textit{Bottom panel (``Case C''):} the same jet is now injected at the centre of a rich galaxy cluster (location marked with `$\bigstar$'). 
The stars (also) mark the centres of the ambient media, while white contours denote pressure isocontours ($\log P \rm{[dyn/cm^2]}$). The indicated $R_{virial}$, $R_{500}$ and $R_{200}$ (radii to over-densities) characterize the ambient environments, and the insets show the corresponding 1D (ambient) profiles of $\rho$ and $P$ along the jet flow direction from their injection location (as marked with `$\boldsymbol{\times}$' for the group and `$\bigstar$' for the clusters).}
\label{Fig:Sim_setup}
\end{figure*}

\begin{table*}[htbp]
\centering
\begin{threeparttable}
\caption{Key jet and environment parameters for the simulations.}
\label{Tab:Simulation_setup}
\begin{tabular*}{\textwidth}{@{\extracolsep{\fill}}lccc}
\hline
\hline
\textbf{Parameters} & \textbf{``Case A''} & \textbf{``Case B''} & \textbf{``Case C''} \\
\hline

\multicolumn{4}{l}{\textup{[Jet properties]}} \\
\hline
Jet number density [cm$^{-3}$] & $5 \times 10^{-8}$ & $5 \times 10^{-8}$ & $5 \times 10^{-8}$ \\
Jet power [erg s$^{-1}$] & $3.5 \times 10^{46}$ & $3.5 \times 10^{46}$ & $3.5 \times 10^{46}$ \\
Jet bulk Lorentz factor at injection, $\Gamma$ & 12 & 12 & 12 \\
Jet magnetization, $\sigma$ & 0.1 & 0.1 & 0.1 \\

\hline
\multicolumn{4}{l}{\textup{[Ambient properties]}} \\
\hline
Central number density [cm$^{-3}$] & $10^{-3}$ & $3 \times 10^{-3}$ & $3 \times 10^{-2}$ \\
Temperature [keV] & $1.1$ & $6.6$ & $7.7$ \\
King model core radius [kpc] & 33 & 125 & 33 \\
King model exponent ($\beta$) & 0.55 & 0.65 & 0.55 \\
Overdensity radii ($R_{500}$, $R_{200}$, $R_{100}$) [Mpc] & (0.45, 0.63, 0.81) & (1.70, 1.21, 2.22) & (1.67, 2.39, 3.18) \\

\hline
\multicolumn{4}{l}{\textup{[Computational domain]}} \\
\hline
$x$ [Mpc] & $[-0.01,\,5.01]$ & $[-0.01,\,3.51]$ & $[-0.01,\,3.51]$ \\
$y,\,z$ [Mpc] & $[-0.3,\,0.3]$ & $[-0.3,\,0.3]$ & $[-0.3,\,0.3]$ \\
Grid size [$x,\,y,\,z$] & $[12550,\;1500,\;1500]$ & $[8800,\;1500,\;1500]$ & $[8800,\;1500,\;1500]$ \\

\hline
\hline
\end{tabular*}
\begin{tablenotes}
\small
\item \textbf{Notes.} We adopt a uniform resolution of five grid cells per jet radius (2 kpc) in all the cases, placing these simulations among the most computationally demanding in this regime. Refer to Section~\ref{Sec:Numerical setup} for a complete description.
\end{tablenotes}
\end{threeparttable}
\end{table*}
We aimed to investigate the propagation limits of extragalactic relativistic jets under conditions where the factors critical to extensive growth
\citep[cf.,][]{Dabhade2023} act in tandem throughout their evolution. To this end, we adopt a fiducial jet model characterized by a bulk Lorentz factor at injection of $\Gamma = 12$, a jet particle density of $\rho_j = 5 \times 10^{-8} \rm{amu\, cm}^{-3}$, and a magnetization parameter $\sigma = 0.1$, defined as the ratio of Poynting flux ($B_j^2/4\pi$) to matter energy flux ($\Gamma^2\rho_j h_j$; $h_j$ being the enthalpy) \citep{Giri2026}. The jet radius is set to $2$ kpc (resolved with 5 grid cells). These parameters correspond to a standard powerful Fanaroff–Riley type II \citep[FRII;][]{Fanaroff1974,Daly2012} RG, as the majority of GRGs belong to this class \citep[e.g.,][]{Dabhade2020_SAGAN,Simonte2024}. The adopted Lorentz factor is also reasonable, as values up to $\Gamma \sim 10$ are usually inferred for classical radio sources \citep{Giovannini2001,Hardcastle2026} and GRGs \citep{Laing2015}, with values reported to reach $\sim 15$ even in several low-power GRGs \citep{Giovannini1999}. This jet configuration yields a jet power of $3.5 \times 10^{46}\ \mathrm{erg\ s^{-1}}$, which we adopt as our reference 
``Case A'', propagating in a galaxy group environment (discussed below). This setup is inspired by recent studies of multi-megaparsec jet propagation \citep{Giri2026_5Mpc}, which explored jet parameters that could generate large-scale outflows as part of our GRG simulation campaign. To 
investigate such a jet's propagation in denser cosmic environments, we perform a pair of additional simulations here, regarded as Cases ``B'' and ``C'', 
of the same jet injected at the centre of different clusters. These configurations are illustrated in Fig.~\ref{Fig:Sim_setup}.

The group medium is modeled as a three-dimensional spherically symmetric atmosphere \citep[using a King $\beta-$density profile;][]{Cavaliere1976} with a central density of $10^{-3}\ \mathrm{amu\ cm^{-3}}$,  a core radius of $33$ kpc, and $\beta =0.55$, in hydrostatic equilibrium, resulting in an isothermal temperature of $kT \sim 1.1\ \mathrm{keV}$ \citep{Giri2025_dynamics}. For this profile, the baryonic mass enclosed within 2 Mpc is $2.8 \times 10^{12} M_{\odot}$, corresponding to a total mass of $\sim 3.2 \times 10^{13}\ M_{\odot}$ assuming a baryon fraction of 0.09 \citep{Paul2017}. This yields a virial radius \citep{Gaspari2020} of $\sim 812\ \mathrm{kpc}$, with $R_{500} \sim 454\ \mathrm{kpc}$, marking the radius up to which the environment is 500 times denser than the critical density of the present day Universe \citep{Sun2009} (similarly, $R_{200} \sim 630\ \mathrm{kpc}$). These values correspond to a standard galaxy group environment relevant to Mpc-scaled GRG jets \citep[see e.g.,][]{Giacintucci2011,Ineson2015,Malarecki2015,Cantwell2020}. 
To be consistent with the environment typically associated with multi-Mpc-scaled GRGs extending beyond $\sim 5$ Mpc \citep{Oei2022_5Mpc,Oei2024_7Mpc}, we need to allow a jet to flow in a very low density ambient medium. So, our reference jet is injected at a distance of $600$ kpc from the center of the galaxy group in ``Case A'', following \cite{Giri2026,Giri2026_5Mpc}. Central AGN jets in virialized environments are typically considered located within $R_{500}$; e.g., \citet{Ma2013} considers sources within $\sim250$~kpc of the halo centre. So, our jet is injected at $600$~kpc, placing it beyond $R_{500}$ but within the virial radius, representing jet propagation through the outer halo environment \citep[also, following][]{Giri2026_5Mpc}.

To assess the robustness of our model against environmental variations, we consider ``Case B'', in which the same jet 
is now propagating in a denser medium representative of a typical galaxy cluster. This environment is characterized by a central density of $0.003\ \mathrm{amu\ cm^{-3}}$ (core radius set at $125$ kpc, with $\beta = 0.65$) and an isothermal temperature of $kT \sim 6.6\ \mathrm{keV}$ \citep{Chen2007}, with pressure and gravitational profiles maintaining hydrostatic equilibrium. In this case, the total mass enclosed within 3.5 Mpc is $6.3 \times 10^{14}\ M_{\odot}$ \citep{Lovisari2020}, corresponding to a virial radius of $\sim 2.22\ \mathrm{Mpc}$ ($R_{500} \sim 1.21$ Mpc), assuming a baryon fraction of 0.14 \citep{Paul2017} (similarly, $R_{200} \sim 1.7\ \mathrm{Mpc}$). 

To understand 
the rarity of extreme GRGs, we performed a third simulation 
in which the same jet propagates through a dense galaxy cluster environment (``Case C''). This ambient medium is modeled with a central density of $0.03\ \mathrm{amu\ cm^{-3}}$, a core radius of $33$ kpc, and $\beta = 0.55$. These parameters correspond to a gravitationally bound cluster atmosphere with a temperature of $\sim 7.7$ keV. Assuming a baryon fraction of $0.14$ (as above), the total mass enclosed within $3.5$ Mpc is approximately 
$1.8 \times 10^{15}\ M_{\odot}$ \citep{Hardcastle2018,Lovisari2020}, yielding a virial radius of $\sim 3.18$ Mpc, with corresponding characteristic radii of $R_{500} \sim 1.67$ Mpc and $R_{200} \sim 2.39$ Mpc. Here, the jets are injected from the centre of the cluster environments, consistent with AGN feedback models in which powerful jets are expected to originate predominantly from central galaxies \citep{Cavagnolo2008}. The jet and ambient medium parameters adopted for the three simulations 
are summarized in Table~\ref{Tab:Simulation_setup}, and the ambient densities are plotted in Fig.~\ref{Fig:Sim_setup}.

We further model the synchrotron emissivity ($\mathcal{J}_{\rm syn}$) arising from non-thermal electrons at an observing frequency of 144 MHz, in order to obtain a realistic representation of the radiative appearance of these jetted systems in the low-frequency radio band probed by instruments such as LOFAR \citep{vanHaarlem2013}, which are particularly effective in discovering GRGs extending beyond $\sim 5$ Mpc \citep{Andernach2025}. The emissivity is calculated following the prescription of \citet{Meenakshi2023}, which has been successfully applied to GRG systems \citep{Giri2025_emission}, as well as on RGs on varying jet power and propagation scales \citep{Meenakshi2024,Giri2026_RGs}. The emissivity is expressed as,
\begin{equation}\label{Eq:Sync_emissivity}
\begin{split}
\mathcal{J}'_{\rm syn} ={} & N_0 \frac{3^{p/2} e^2 \nu'^{-(p-1)/2} |\mathbf{B'} \times \mathbf{\hat{n}}'|^{(p+1)/2}}{2c(p+1)} 
\mathcal{G}\left( \frac{p}{4} + \frac{19}{12} \right) \\
& \times \mathcal{G}\left( \frac{p}{4} - \frac{1}{12} \right)\left( \frac{e}{2\pi m_e c} \right)^{(p+1)/2}~,
\end{split}
\end{equation}
for a population of non-thermal electrons (mass $m_e$, charge $e$) described by a power-law energy distribution with slope $p = 2.2$, spanning Lorentz factors in the range $10^2 \leq \gamma \leq 10^6$. Here, $c$ denotes the speed of light, $\mathbf{B^\prime}$ denotes the magnetic field in the comoving frame of the fluid, $\mathbf{\hat{n^\prime}}$ represents the line-of-sight direction in the comoving frame, and $\mathcal{G}$ is the Gamma function. The normalization constant $N_0$ defines the amplitude of the particle energy distribution and is determined by assuming that a fraction $\epsilon$ of the local thermal energy density ($E_{\mathrm{th}}$) is transferred to the non-thermal particle population. Accordingly, $N_0$ is given by,
\begin{equation}
N_0 = \frac{\epsilon E_{\rm th}}{m_e c^2 \int_{10^2}^{10^6} \gamma^{-p+1}d\gamma}.
\end{equation}
We adopt $\epsilon = 0.5$, in line with values that have been shown to reproduce the observed emission properties of GRGs as well as smaller-scale radio jets \citep{Giri2025_emission,Giri2026_RGs}. Quantities denoted with primes in Eq.~\ref{Eq:Sync_emissivity} refer to the comoving frame of the fluid and are transformed to the observer frame via,
\begin{equation}\label{Eq:Synchrotron_transforms}
\begin{split}
\mathcal{J}_{\rm syn} (\nu, \mathbf{\hat{n}}, \mathbf{B}) &= \mathcal{D}^2 \mathcal{J}'_{\rm syn} (\nu', \mathbf{\hat{n}'}, \mathbf{B}'); \\[4pt]
\nu' &= \nu/\mathcal{D}; \\[4pt]
\mathbf{\hat{n}}' &= \mathcal{D}\left[\mathbf{\hat{n}}+\left(\frac{\gamma^2}{\gamma + 1}\mathbf{\beta} \cdot \mathbf{\hat{n}} -\gamma \right) \mathbf{\beta} \right]; \\[4pt]
\mathbf{B}' &= \frac{1}{\gamma}\left[\mathbf{B}+ \frac{\gamma^2}{\gamma + 1}(\mathbf{\beta}\cdot \mathbf{B})\mathbf{\beta} \right]
\end{split}
\end{equation}
where, $\mathbf{B}$ denotes the magnetic field in the observer frame, $\mathcal{D}$ is the Doppler factor, $\nu = 144$ MHz is the observing frequency, and $\mathbf{\hat{n}}$ represents the line-of-sight direction in the observer frame, taken to be along the $z$-axis. 
This choice corresponds to viewing the system in the plane of the sky, as is often essentially the case for GRGs \citep{Oei2023}.
Finally, the observable intensity map ($\mathcal{I}_{\rm 144\, MHz}$) is obtained by integrating the emissivity ($\mathcal{J}^{\rm 144\, MHz}_{\rm syn}$) along the line of sight (the $z$-axis).

%
\section{Results}\label{Sec:Results}
%

We carried out three high-resolution simulations to investigate the propagation limits of extragalactic jets and to assess the robustness of these flows. The  
parameters of these cases are summarized in Table~\ref{Tab:Simulation_setup} (also, Fig.~\ref{Fig:Sim_setup}). These simulations rank among the most computationally demanding studies of extragalactic jet propagation performed in a zoom-in framework (as opposed to fully cosmological simulations). Consequently,  our initial exploration of 
parameter space has had to be
limited to these cases, which should be considered as a 
demonstration that relatively young jets can reach multi-Mpc scales.

\subsection{``Case A'': a jet pair reaching 9 Mpc emerging from a galaxy group}
\begin{figure*}[h]
\centering
\includegraphics[width=\textwidth]{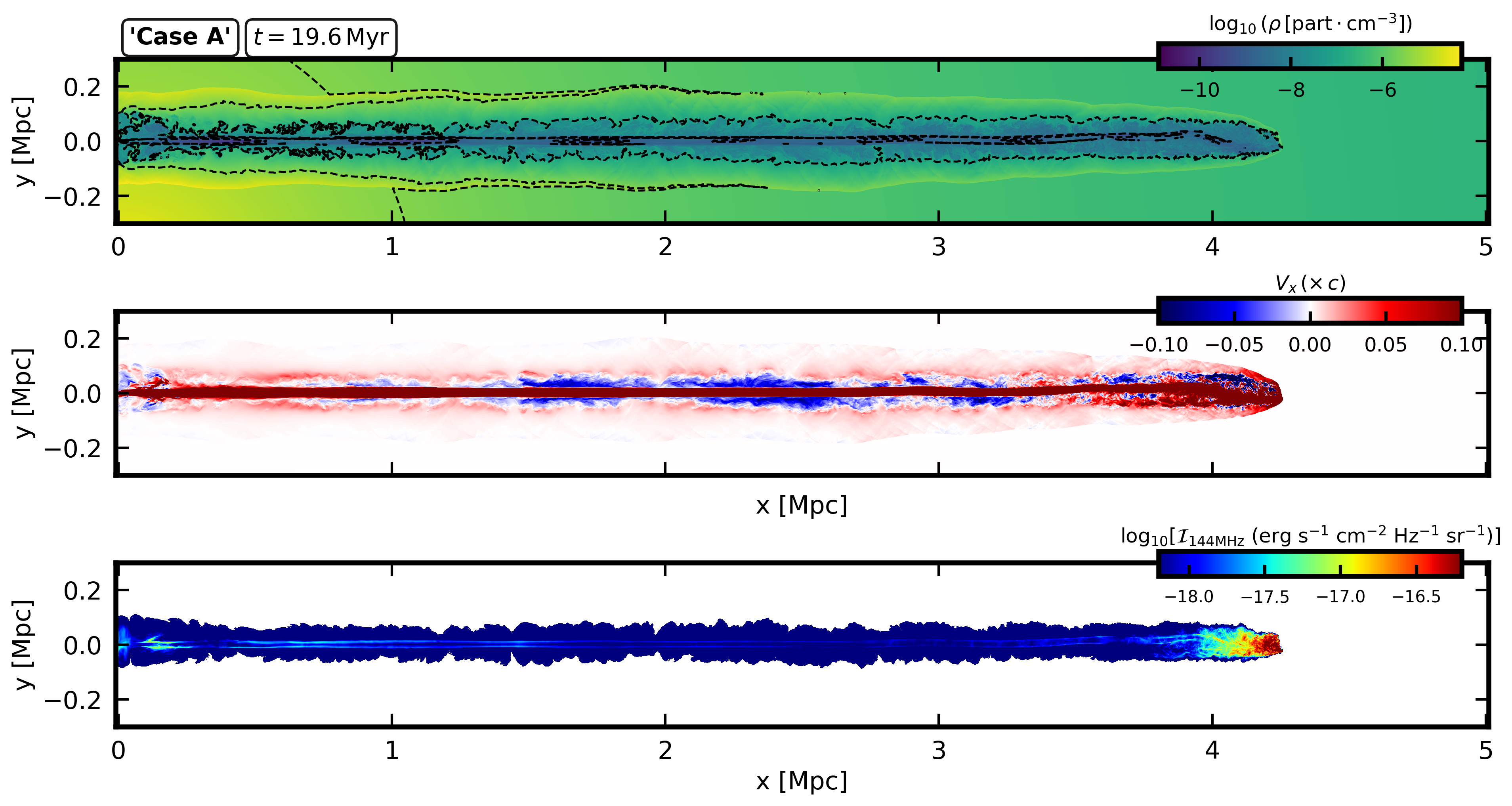}
\caption{\textbf{``Case A'':} Jet with kinetic power of $3.5 \times 10^{46}$ erg s$^{-1}$, reaching a one-sided extent of $\sim 4.25$ Mpc in 19.6 Myr (evolution in galaxy group, subsequently in cosmic void). This demonstrates the physical plausibility of extragalactic jetted systems exceeding the currently observed maximum total extents of $\sim 7$ Mpc. \textit{Top}: density slice illustrating the jet–environment interaction (contours: log$\rho \equiv -9,-7,-5.5$); \textit{middle}: $x-$velocity ($V_{x}$) slice highlighting the jet’s forward advance (in red) and associated backflow (in blue) (colorbar is truncated for better visualization); \textit{bottom}: line-of-sight integrated intensity ($\mathcal{I}_{\rm 144\, MHz}$) tracing enhanced emission near the jet base as well as the terminal hotspot.}
\label{Fig:8Mpc_collage}
\end{figure*}

Our reference ``Case A'' demonstrates efficient large-scale jet propagation, reaching a (one-sided) 
length of $\sim 4.25$ Mpc within $19.6$ Myr, corresponding to an average advance speed of $0.72 c$. The density structure, shown in Fig.~\ref{Fig:8Mpc_collage} (\textit{top}), reveals rapid jet advancement indicated by the continued 
presence of a bow shock, shocked rim, and a laterally confined cocoon with a high axial ratio (length-to-width $\sim 20$), consistent with the morphology of high-power large-scale flows \citep[e.g.,][]{Machalski2008,Sebastian2018,Oei2024_7Mpc}. The overlaid contours highlight the jet’s ability to drill through the group medium and subsequently evolve into the lower-density cosmic void environment. 

To further characterize the cocoon dynamics, we examine the velocity field along the jet axis ($V_x$). The onset of backflows is clearly observed near the jet head, where jet–environment interactions are strongest. Owing to the high jet velocity and strong interaction-driven entropy gradients \citep{Cielo2017}, the backflow speeds become comparable to the forward flow, leading to the formation of a 
thin cocoon. 
Note that the presented maps correspond to 2D slices ($x–y$ plane; $z=0$; Fig.~\ref{Fig:8Mpc_collage}), illustrating 
the complex backflow pattern present in the tri-axial cocoon.

We assess the observational relevance of our simulated structures by computing an approximate synchrotron emission map at 144 MHz ($\mathcal{I}_{\rm 144\, MHz}$). Enhanced emission is observed near the base of the jet in the recollimation zones, as well as near the jet's head. These features are physically well-motivated as both the recollimation and termination shocks locally amplify thermal energy, magnetic fields and particle densities, 
as further discussed in Section.~\ref{Sec:Evolution of the jet(s)}. The resulting emission morphology (enhanced emission near the jet's base, followed by relatively low emission along most of the jet's extent and thereby hotspots at jet termini) is particularly consistent with that inferred for recently observed $\sim 7$ Mpc-scale jets \citep{Oei2024_7Mpc} (refer to also, Fig.~\ref{Fig:Porphyron}). 
The integrated flux density of the (one-sided) simulated system is $39$ mJy, consistent with values reported for Porphyrion at $z = 0.896$.

We note that ``Porphyrion'' (Fig.~\ref{Fig:Porphyron}, for a reference) exhibits a prominent inner hotspot.  
A similar, but somewhat weaker, feature in our simulation, would be enhanced if particle acceleration processes were explicitly modeled in our emission calculations \citep{Kundu2022,Dubey2024}. However, incorporating such physics at the required resolution remains computationally prohibitive.  
Consequently, we cannot  determine whether the observed inner hotspot is primarily associated with enhanced emission due to recollimation-driven compression \citep[in GRGs, e.g.,][]{Saripalli1994,Safouris2009}, or if it instead traces the hotspot of a restarted jet \citep[in GRGs, e.g.,][]{Dabhade2025}.

Despite propagating to extreme distances, the jet remains largely stable against disruptive magneto-hydrodynamical instabilities and ambient matter entrainment, although mild lateral bending near the jet head indicates the onset of weak decollimation. Overall, the simulated system reproduces radio sources extending to $\sim 9$ Mpc and suggests that even larger 
systems are physically plausible. So our theoretical results provide a predictive framework for upcoming observations, particularly with the latest LOFAR survey data releases \citep{Shimwell2026} and the future Square Kilometre Array \citep{Kapinska2015}. 
 
\subsection{Cases ``B'' \& ``C'': jet pairs stalling around 3 Mpc in clusters}

To assess whether large-scale jet propagation can be sustained in denser cosmic environments, we performed additional simulations, referred to as ``Case B'' and ``Case C'', in which the same jet as in ``Case A'' is injected within the core of a typical galaxy cluster and that of a denser cluster, respectively. The properties of these jet–environment systems are  
in Table~\ref{Tab:Simulation_setup} (also see, Fig.~\ref{Fig:Sim_setup}).

\begin{figure*}[h]
\centering
\includegraphics[width=\textwidth]{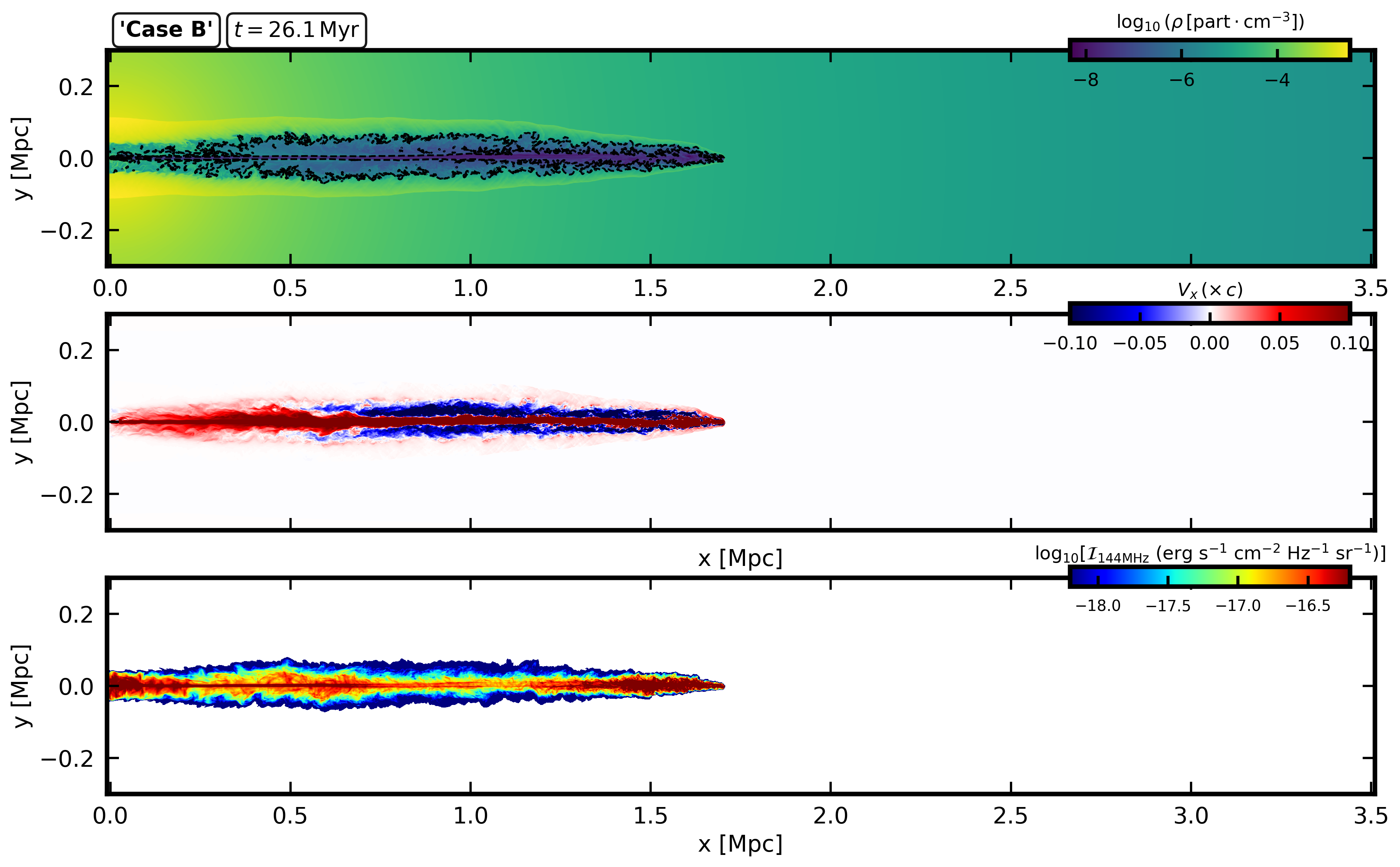}
\caption{\textbf{``Case B'':} Jet with a kinetic power of $3.5 \times 10^{46}$~erg~s$^{-1}$, reaching a one-sided extent of $\sim1.7$~Mpc after 26.1~Myr. The denser ambient medium corresponding to a typical cluster, results in a slower advance speed of the jet head compared to ``Case A'', while maintaining a collimated large-scale morphology. \textit{Top:} gas density distribution. \textit{Middle:} x-component of the velocity field. \textit{Bottom:} line-of-sight integrated synchrotron intensity map at 144~MHz.}
\label{Fig:Collage_Typical_Cluster}
\end{figure*}

As shown in Fig.~\ref{Fig:Collage_Typical_Cluster} for ``Case B'', the jet exhibits a rapidly propagating, arrowhead-shaped cocoon morphology. Because of the greater 
density of the ambient medium its advance speed is lower than in ``Case~A'', and a 
one-sided extent of $\sim1.7$~Mpc is reached after 26.1~Myr, corresponding to an average expansion speed of $\sim0.22c$ 
and a length-to-width axial ratio of 14.3. Although the jet remains collimated, a gradual widening of the beam is apparent, indicating the onset of the decollimation phase. A classical backflow pattern is also observed, with jet material streaming back from the hotspot toward the source core. The denser environment, however, impedes the backflow, preventing it from efficiently reaching the jet base and consequently beginning to form a broader cocoon structure. 

The emission morphology is particularly noteworthy, displaying a network of internal filamentary features suggestive of (magnetized) turbulent motions within the cocoon plasma \citep[e.g.,][]{Giri2025_emission}. In addition to the enhanced emission at the terminal hotspot, a distinct region of elevated emissivity is present near the midsection of the source. Combined, these features are reminiscent of
a restarted radio source, despite originating solely from the internal dynamics of a continuously powered jet \citep{Young2025,Giri2025_emission}. 

\begin{figure*}[h]
\centering
\includegraphics[width=\textwidth]{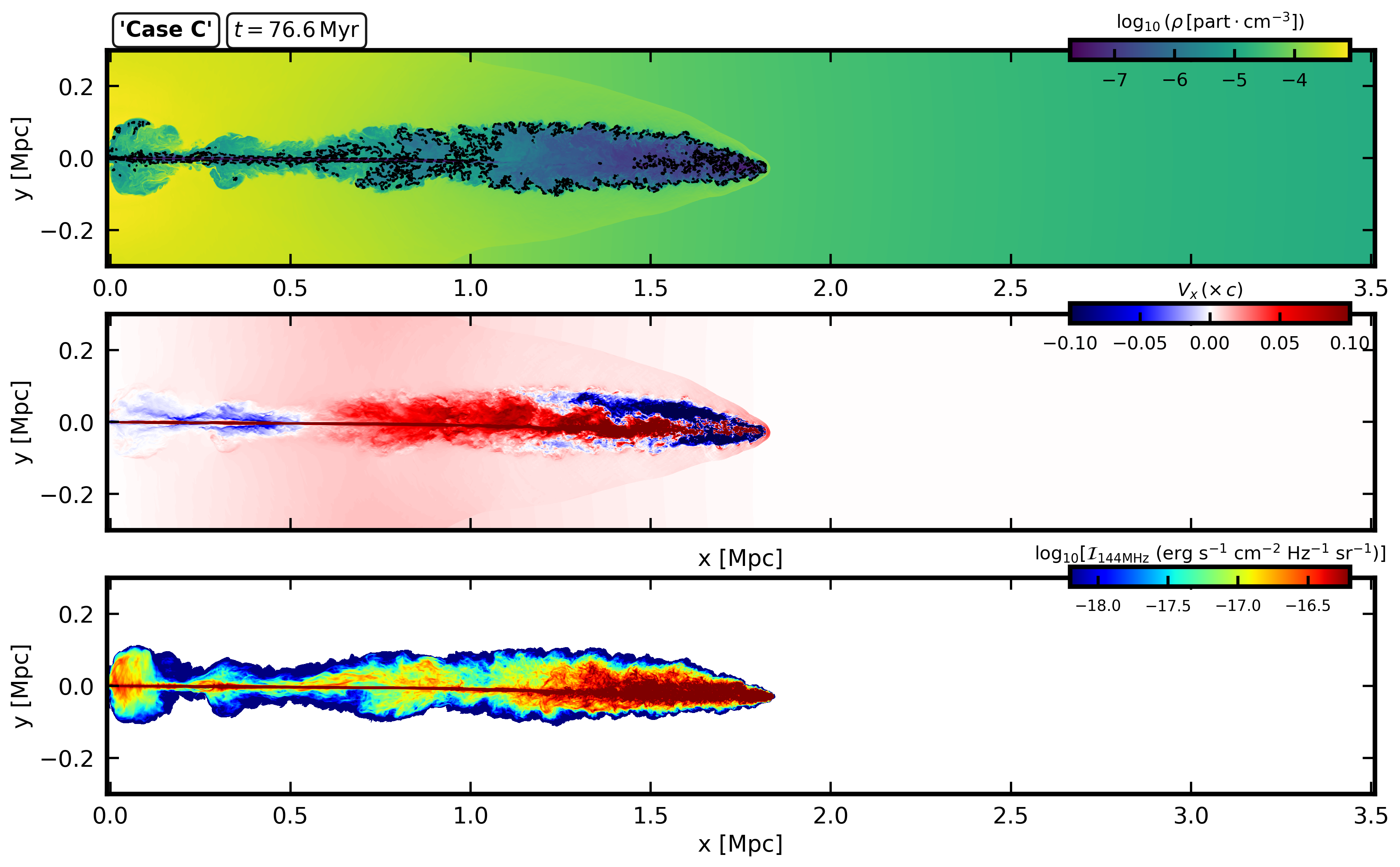}
\caption{\textbf{``Case C'':} Jet with kinetic power of $3.5 \times 10^{46}$ erg s$^{-1}$, evolving within a rich cluster potential, reaching a one-sided extent of $\sim 1.9$ Mpc after 76.6 Myr. This evolution illustrates why extremely large GRGs (sizes exceeding $\sim 5$ Mpc) are rare in dense clusters, while sources stalling around $\sim 3$ Mpc are comparatively common in such environments. The panels show slices of the gas density (\textit{top}), the $x-$component of the velocity (\textit{middle}), and the line-of-sight integrated intensity at 144 MHz (\textit{bottom}), tracing jet dynamics and associated radiative signatures.}
\label{Fig:Cluster_collage}
\end{figure*}

In ``Case C'', the jet undergoes significant deceleration, reaching a (one-sided) distance of $\sim 1.9$ Mpc after $76.6$ Myr  (Fig.~\ref{Fig:Cluster_collage}), yielding an average head speed of $\sim 0.08c$; similar to e.g., \citet{Machalski2008}. The density structure reveals the emergence of a wider lobe, and more jet decollimation, with a lower axial ratio of 8.6 for the cocoon. The corresponding velocity field further highlights enhanced backflows incorporating complex three-dimensional flows, with prominent negative velocity regions (blue in Fig.~\ref{Fig:Cluster_collage} \textit{middle}) tracing the flow of plasma back towards the injection region. These flows reflect stronger jet–environment coupling and more efficient energy dissipation within the dense cluster medium, as further evidenced by the presence of a pronounced and widened bow shock. The jet is clearly seen to begin to wiggle at and beyond $\sim 1$ Mpc, as is a $\sim 2^{\circ}$ global bend in the jet propagation, marking the beginning of jet decollimation. The intensity map ($\mathcal{I}_{\rm 144\, MHz}$; Fig.~\ref{Fig:Cluster_collage}, \textit{bottom}) shows a clear concentration of emission in the lobes. The observational relevance of such systems is discussed later (see, Fig.~\ref{Fig:GRG-BCGs}, for reference).

\subsection{Comparing the cases}

\begin{figure}[h]
\centering
\includegraphics[width=\columnwidth]{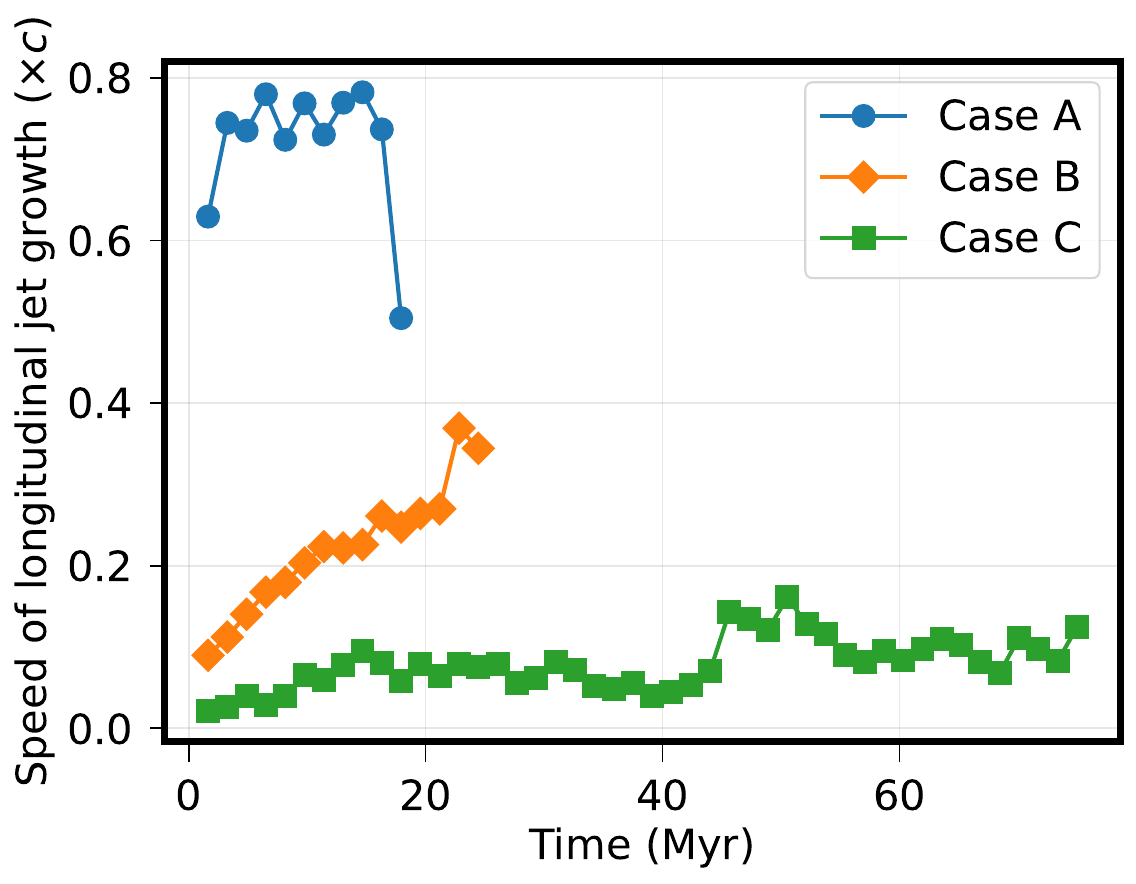}
\caption{Longitudinal expansion speed of the jetted structures for the three simulation cases considered in this work. The figure illustrates the greater propensity for extreme giant radio galaxies to develop in relatively underdense environments, where reduced confinement enables sustained rapid growth. The advance speed eventually flattens or declines following the onset of jet decollimation and the development of MHD instabilities.}
\label{Fig:Jet_expansion_speed}
\end{figure}

When comparing the linear expansion speed of the jet in the three ambient environments, the clear differences in their evolutionary histories are apparent in Fig.~\ref{Fig:Jet_expansion_speed}. In all cases, the jet head initially propagates at a relatively low speed thorough the denser central regions 
of the host environment. As the jet advances into progressively rarefied regions, its expansion speed increases in response to the declining ambient density. 
This acceleration phase does not continue indefinitely; the propagation speed eventually stabilizes or decreases as the jet begins to undergo undulation and bending that can lead to decollimation, marking the transition from a well-collimated to a more disrupted flow.
Fig.~\ref{Fig:Jet_expansion_speed} clearly illustrates why the formation of GRGs of extreme scales is more favorable in group-like environments \citep[or generally in low-dense cosmic locations;][]{Mahato2026} than in clusters. The average expansion speed of $0.72c$ for ``Case A'' is high but has been found in numerical simulations where jets are set to propagate through a diluted environment \citep{Walg2020,Ratnaparkhi2026}. In ``Case A'', the very weak environmental confinement once in the void region allows the jet to maintain a high advance speed, facilitating the growth of multi-Mpc-scale structures. However, ``Case B'' and ``Case C'', owing to their denser environments, undergo a considerable reduction in propagation speed for identical jet powers, causing the structures to stall at one-sided scales of approximately $1.5\text{--}2\text{ Mpc}$. These results are consistent with recent observational findings \citep{Sankhyayan2024,Andernach2025} and support the observational inferences discussed in Section~\ref{Sec:A comparison with observations}.

We also examine the energy budget injected by these powerful jets, which maintain a constant jet power across cases. We emphasize that the total energy deposited (mechanical plus thermal) reaches [$0.9,\, 1.2,\, 3.4$] $\times 10^{62}\text{ erg}$ for Cases ``A'', ``B'', and ``C'' at ages [$19.6,\, 26.1,\, 76.6$] Myr, respectively (one-sided). This energetic output is predominantly deposited within $R_{500}$ of the host clusters (in ``B'', ``C''), establishing GRGs as crucial sources of thermalization in intra-cluster media. Conversely, extreme-scale GRGs that expand beyond group-scale boundaries (in ``A'') act as prominent feedback conduits, driving thermal, mechanical, and magnetic transport directly into the surrounding cosmic voids \citep[see, also,][]{Giri2026_5Mpc}.

\subsection{Evolution of the jet(s)} \label{Sec:Evolution of the jet(s)}

\begin{figure*}[h]
\centering
\includegraphics[width=\textwidth]{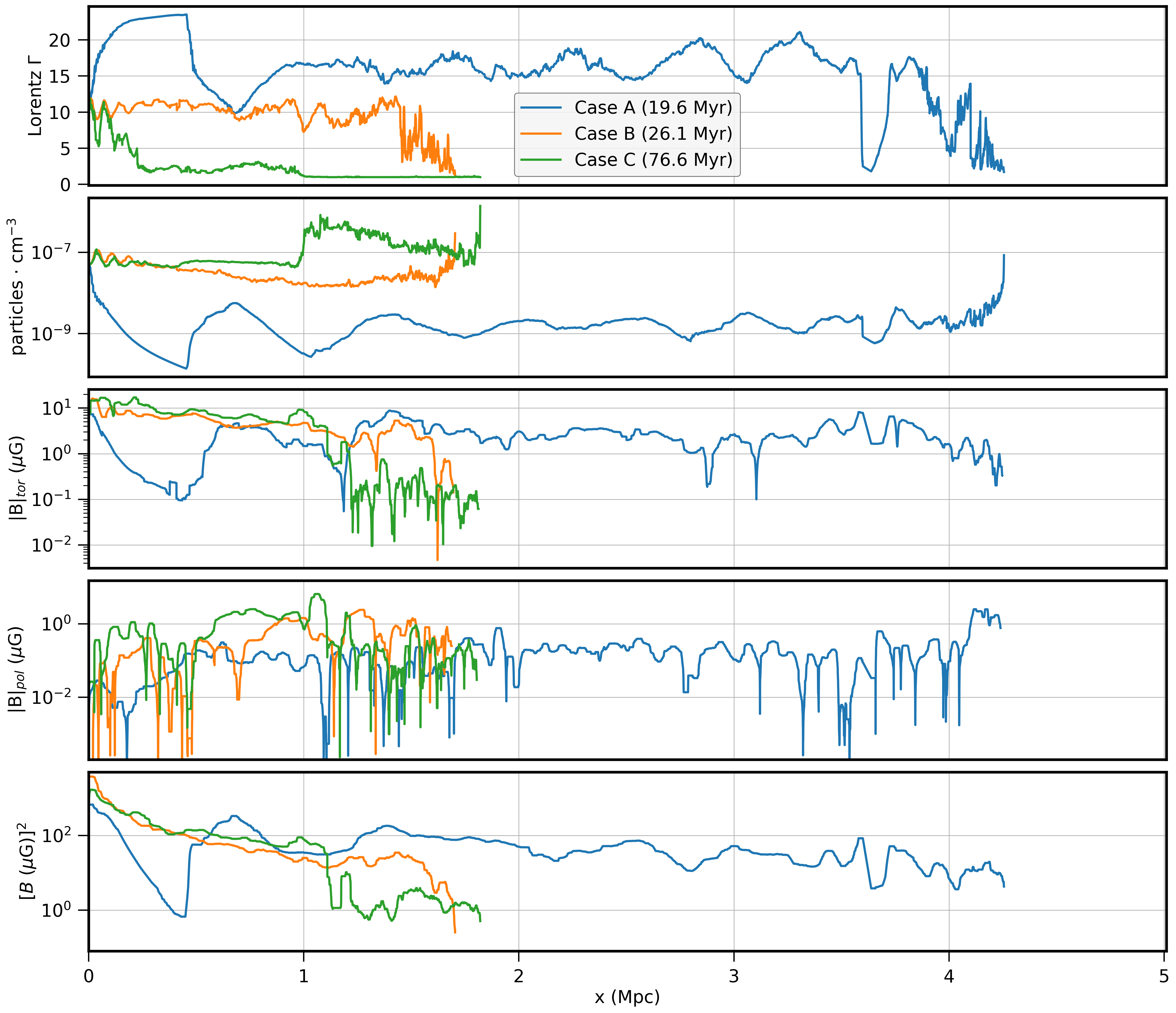}
\caption{Plots of the longitudinal (jet-spine) properties of the evolved jets for the three cases considered. From top to bottom the panels illustrate: the jet beam's motion (expressed via Lorentz factor); particle number density; and the toroidal, poloidal, and the total magnetic field components. The comparisons highlight the transition between collimated and decollimated flow regions in different contrasting environmental conditions, given that the injected jets are the same in all cases, demonstrating how external medium properties influence jet propagation and structural stability.}
\label{Fig:1D_variation}
\end{figure*}

In Fig.~\ref{Fig:1D_variation}, we present the longitudinal variation of key jet properties along the jet beam (spine) to diagnose the internal dynamics of the largest simulated jets and provide insights into the physical processes operating in GRGs.

For ``Case A'', the top panel of Fig.~\ref{Fig:1D_variation} shows the evolution of the Lorentz factor, $\Gamma$. Although the jet is injected with $\Gamma=12$, it is accelerated to values exceeding 20 before returning to the injection level, producing a quasi-periodic oscillatory pattern along the jet. These oscillations mark the locations of successive recollimation shocks \citep{VanDer2019}, while the pronounced decrease near the jet head reflects the strong deceleration due to the interaction with the ambient medium \citep{Rossi2024}.

The first recollimation shock is particularly strong, producing significant compression of the jet plasma and an enhancement of the toroidal magnetic field (and thermal energy density). Consequently, this region is expected to exhibit enhanced synchrotron emission. This interpretation is supported by Fig.~\ref{Fig:Jet_seath_20Myr}, where the brightest inner emission feature spatially coincides with the first recollimation shock. As the jet evolves, both the recollimation shock and the associated emission enhancement shift downstream, demonstrating that the inner brightening is a direct consequence of recollimation-induced compression.

This result provides a plausible explanation for the prominent inner knot observed in Porphyrion (Fig.~\ref{Fig:Porphyron}). Nevertheless, the morphology and brightness of the knot may also be consistent with a restarted jet \citep{Chavan2023,Kumari2026}. In our emission model, accounting for particle acceleration at shocks should further amplify the local synchrotron emissivity \citep{Vaidya2018} at these recollimation sites. Distinguishing between these scenarios therefore requires dedicated simulations of restarted jets incorporating a more realistic treatment of particle acceleration, which makes such projects computationally prohibitive at these spatio-temporal scales.

For the remaining two cases, where the jets propagate through denser ambient environments (in comparison to ``Case A''), dissipative processes become prominent from the early stages of their evolution (Fig.~\ref{Fig:1D_variation}). As a result, the characteristic recollimation pattern is suppressed, leading to a reduced advance speed of the jet head, consistent with the trends shown in Fig.~\ref{Fig:Jet_expansion_speed}. The dissipation is particularly pronounced in ``Case C'', where the jet undergoes substantial decollimation and energy loss.

The evolution of the poloidal magnetic field further indicates that a poloidal component can be generated from an initially purely toroidal magnetic configuration. In well-collimated jets, this field amplification is primarily associated with strong recollimation shocks and the shear layers they produce, whereas in the more dissipative cases it is driven predominantly by enhanced lateral plasma motions accompanying the onset of jet decollimation \citep{Giri2026_5Mpc}.

In Fig.~\ref{Fig:Jet_seath_20Myr}, we present the spatio-temporal evolution only for ``Case A'', our reference model, as it best illustrates the dynamics of extremely large jets such as Porphyrion. The locations of the recollimation shocks are readily identifiable across all panels. These features are already well developed by 9.8~Myr; however, as the jet evolves, the first recollimation shock remains prominent while the downstream shocks progressively weaken, giving rise to a hollow channel along the jet spine. Such behaviour has previously been reported in numerical studies spanning a range of jet scales \citep{Dubey2023,Giri2025_dynamics,Giri2025_dynamics}.

The magnetic field evolution, particularly of its toroidal component, shown in the fourth panel of Fig.~\ref{Fig:1D_variation} reveals localized enhancements at the recollimation shocks and along the jet beam's edges, consistent with earlier studies \citep{Mizuno2015}. The poloidal component appears to be  generated through shock-induced shear layers. The transverse velocity distribution ($V_y$), together with the overlaid contours of the $V_z$ component, further suggests the onset of helicoidal motion within the jet beam (downstream).

Finally, the progressive decline in the synchrotron intensity (Fig.~\ref{Fig:Jet_seath_20Myr}A--C) as the jet evolves reflects the decreasing magnetic energy density and particle energy caused by the expanding cocoon volume. This naturally reduces the surface brightness of the source and provides a plausible explanation for the observational rarity of such extreme GRGs, many of which may evolve below the sensitivity limits of current radio facilities \citep{Koribalski2026,Hardcastle2026}. We reiterate that our simulations do not explicitly track the (spatio-temporal) radiative history of non-thermal electron populations, a computationally challenging task deferred to future work.

\begin{figure*}[h]
\centering
\includegraphics[width=\textwidth]{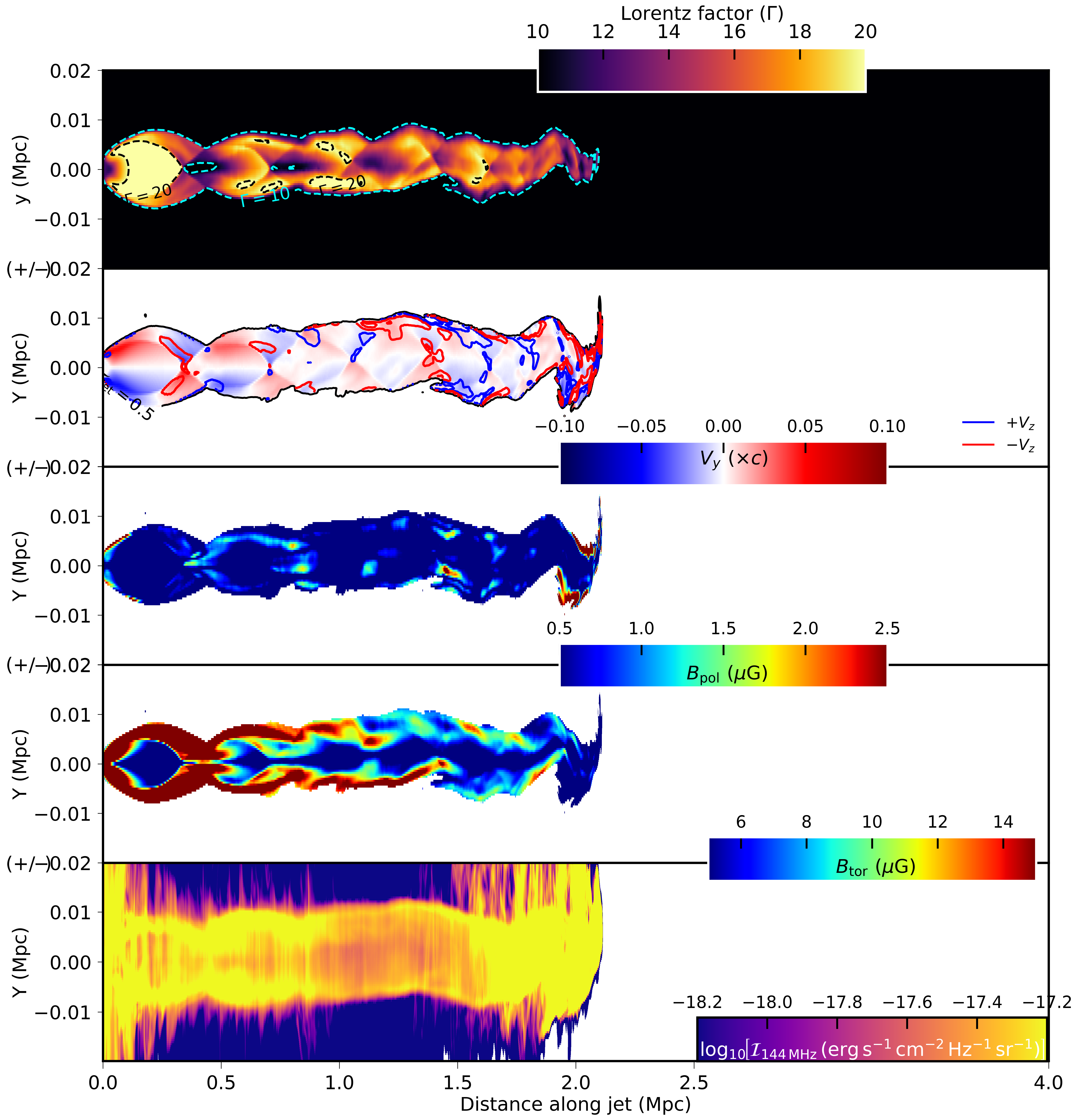}
\caption{\textbf{(A)} Zoomed structural diagnostics of the jet beam of ``Case A'' at 9.8 Myr (\textit{top four panels:} $x-y$ slice at $z=0$; \textit{bottom panel}: line-of-sight integrated map along $z$-axis). \textit{Top:} Lorentz factor distribution with $\Gamma=20$ (in \textit{black}) and $\Gamma=10$ (in \textit{cyan}) contours marking fast and slow propagation boundaries. \textit{Second:} $V_y$ colormap with overlaid $V_z$ contours (\textit{blue}: $+V_z$, \textit{red}: $-V_z$) at $0.004c$ and $0.01c$ magnitude values, indicating the onset of mild helical motion. \textit{Third:} Poloidal magnetic field ($B_{\rm pol}$), tracing internal and boundary stress layer locations of the jet beam.
\textit{Fourth:} Toroidal magnetic field component ($B_{\rm tor}$) is concentrated at the jet edge, providing magnetic tension that aids in limiting lateral dissipation. \textit{Bottom:} projected intensity map showcasing enhanced emission at the jet base (recollimation zones) and head locations.}
\label{Fig:Jet_seath_20Myr}
\end{figure*}

\begin{figure*}[ht]
\ContinuedFloat
\centering
\includegraphics[width=\textwidth]{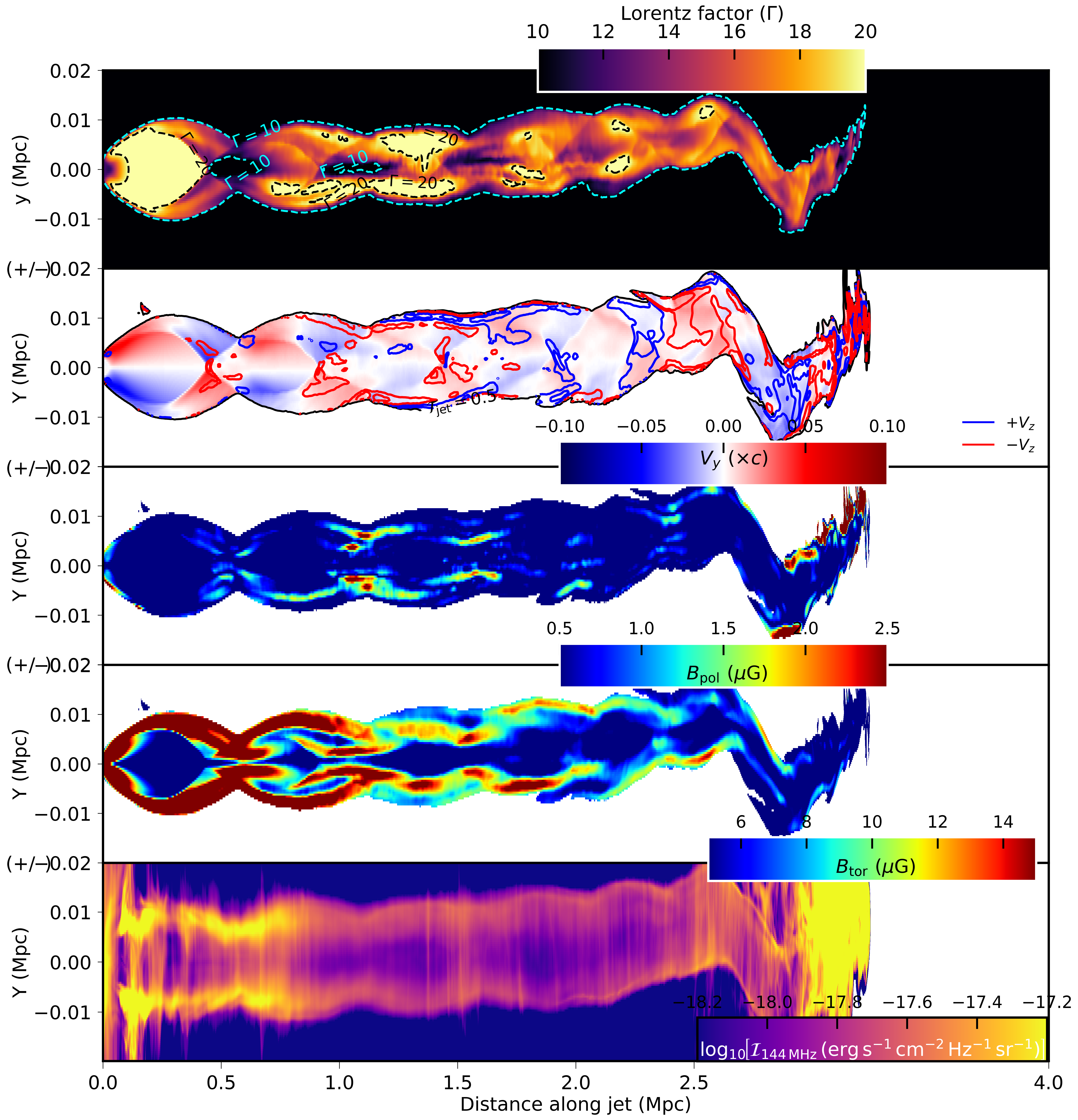}
\caption{\textbf{(B)} Same as Fig.~\ref{Fig:Jet_seath_20Myr}(A), but at 14.7 Myr. The recollimation shock structures are more prominently developed at this evolutionary stage, as evident across the subpanels. Corresponding enhancements in the synchrotron emission are clearly associated with these recollimation zones, particularly at the first recollimation shock where the compression is strongest. In addition, the jet head exhibits the initial signatures of decollimation, indicating the onset of an undulation.}
\end{figure*}

\begin{figure*}[ht]
\ContinuedFloat
\centering
\includegraphics[width=\textwidth]{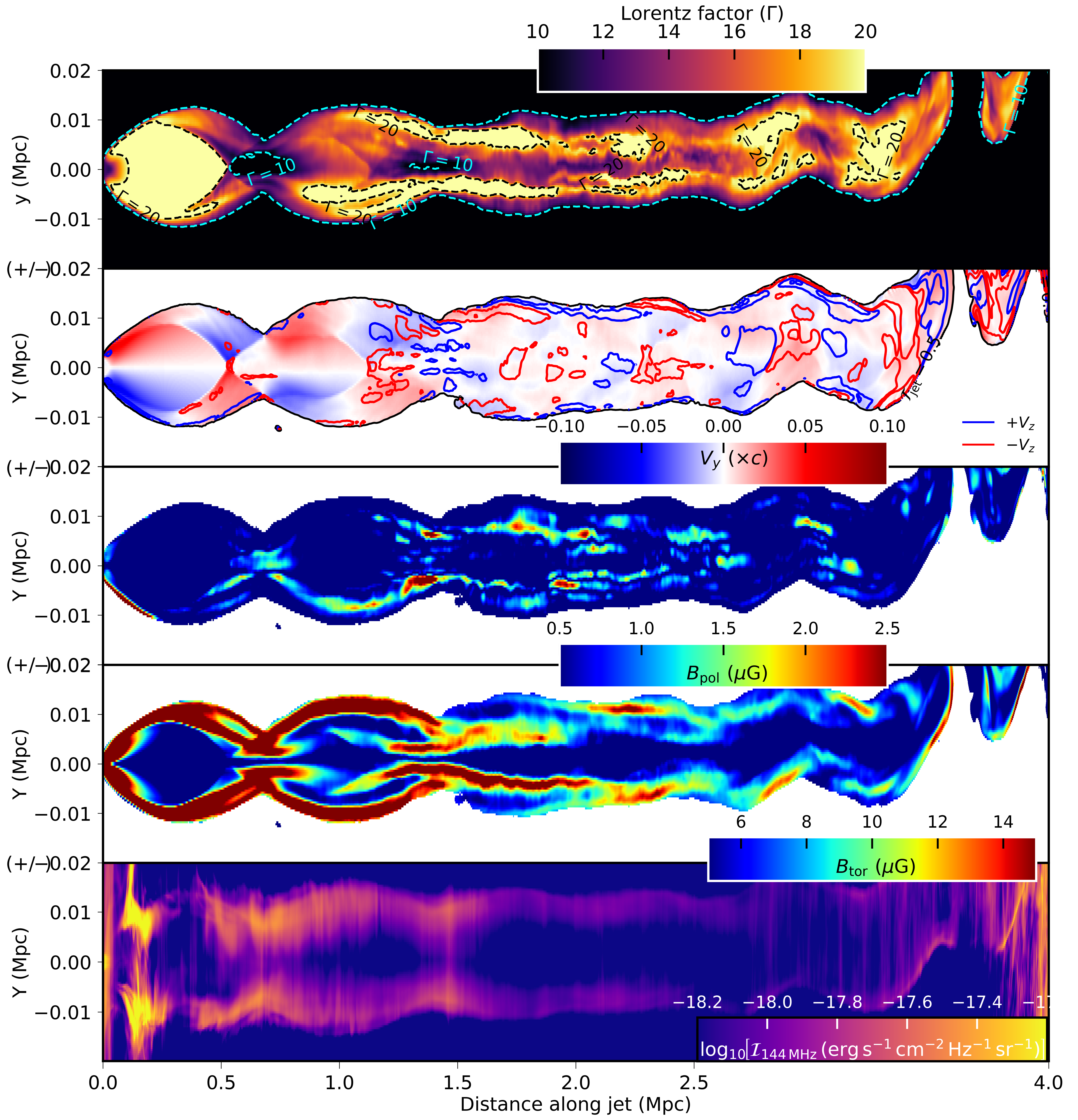}
\caption{\textbf{(C)} Same as Fig.~\ref{Fig:Jet_seath_20Myr}(A), but at 19.6 Myr. The evolution of the recollimation shock, particularly the first recollimation feature, is evident through its increase in both longitudinal and lateral extent. The corresponding shift of the emission region (compared to 9.8 Myr) closely trace the recollimation shock, indicating that the enhanced emission near the jet base (of sources such as Porphyrion) may arises from compression at the recollimation site.}
\end{figure*}

\subsection{Jet inertia against lateral instabilities} \label{Sec:Jet inertia against lateral instabilities}

\begin{figure*}[h]
\centering
\includegraphics[width=\textwidth]{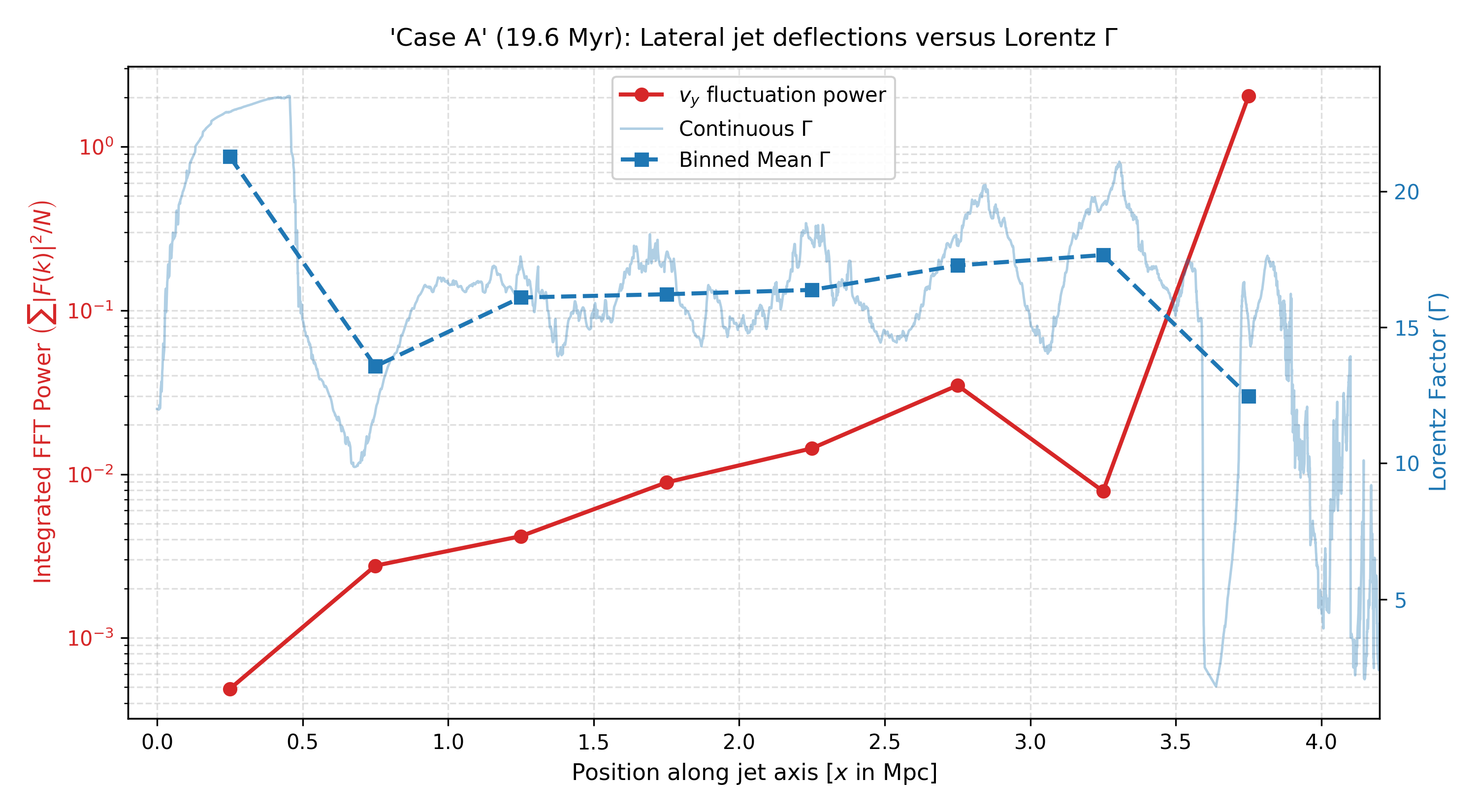}
\caption{Growth of lateral MHD instabilities (as measured along the jet spine) evaluated via $V_y$ FFT power (\textit{left axis}) compared with the jet spine Lorentz factor $\Gamma$ (\textit{right axis}). Binned median $\Gamma$ values (\textit{bold squares}) represent (jet)section-averaged beam speeds alongside the continuous profile (\textit{faded line}; similar to Fig.~\ref{Fig:1D_variation}). The strong suppression of velocity fluctuations at high $\Gamma$ confirms that relativistic thrust stabilizes the jet column against lateral deflection, explaining why extreme GRGs require high jet power to remain collimated over multi-megaparsec scales.}
\label{Fig:FFT_LorG}
\end{figure*}

To understand why extreme GRGs require high jet kinetic powers and to clarify the physical mechanisms governing their rarity, we examine the local relationship between the advance of relativistic beams and the growth of lateral instabilities. Fig.~\ref{Fig:FFT_LorG} illustrates the spatial evolution of the jet spine's Lorentz factor ($\Gamma$) alongside the integrated $V_y$ Fast Fourier Transform (FFT) power ($\sum |F(k)|^2 / N$; $k$ denotes the wavenumber and $N$ is the (grid point) normalization factor within the periodic spatial slabs of the spine (0.5 Mpc) for the FFT) for our benchmark ``Case A'' at $19.6~\text{Myr}$. The FFT power serves as a quantitative metric for non-axisymmetric lateral wiggling (hinting at transverse momentum transport) along the jet column \citep[e.g.,][]{Guan2014,Upreti2024,Giri2026_5Mpc}.

A clear anti-correlation emerges between the beam's relativistic inertia and the amplification of lateral undulations. In the inner regions ($\lesssim 3$ Mpc), where the jet maintains a high bulk Lorentz factor, the growth of lateral fluctuations remains heavily suppressed (amplitude $\lesssim 10^{-2}$). However, as the jet approaches the terminal working surface, mass entrainment, ambient shear-layer interactions, and kinetic dissipation cause $\Gamma$ to drop sharply. This loss of relativistic momentum flux triggers a rapid, order-of-magnitude spike in the lateral fluctuation power, signaling catastrophic lateral decollimation.

This result makes clear the strong interplay between high advance speeds and instability mitigation: high relativistic inertia intrinsically suppresses the growth rates of Kelvin-Helmholtz and current-driven kink instabilities,  \citep[similar to,][]{Mukherjee2020,Giri2026_5Mpc}, thus providing fast propagation and instability suppression simultaneously.

\section{Discussion}\label{Sec:Discussion}

\subsection{Comparisons with observations}\label{Sec:A comparison with observations}

\begin{figure*}[h]
\centering
\includegraphics[width=\textwidth]{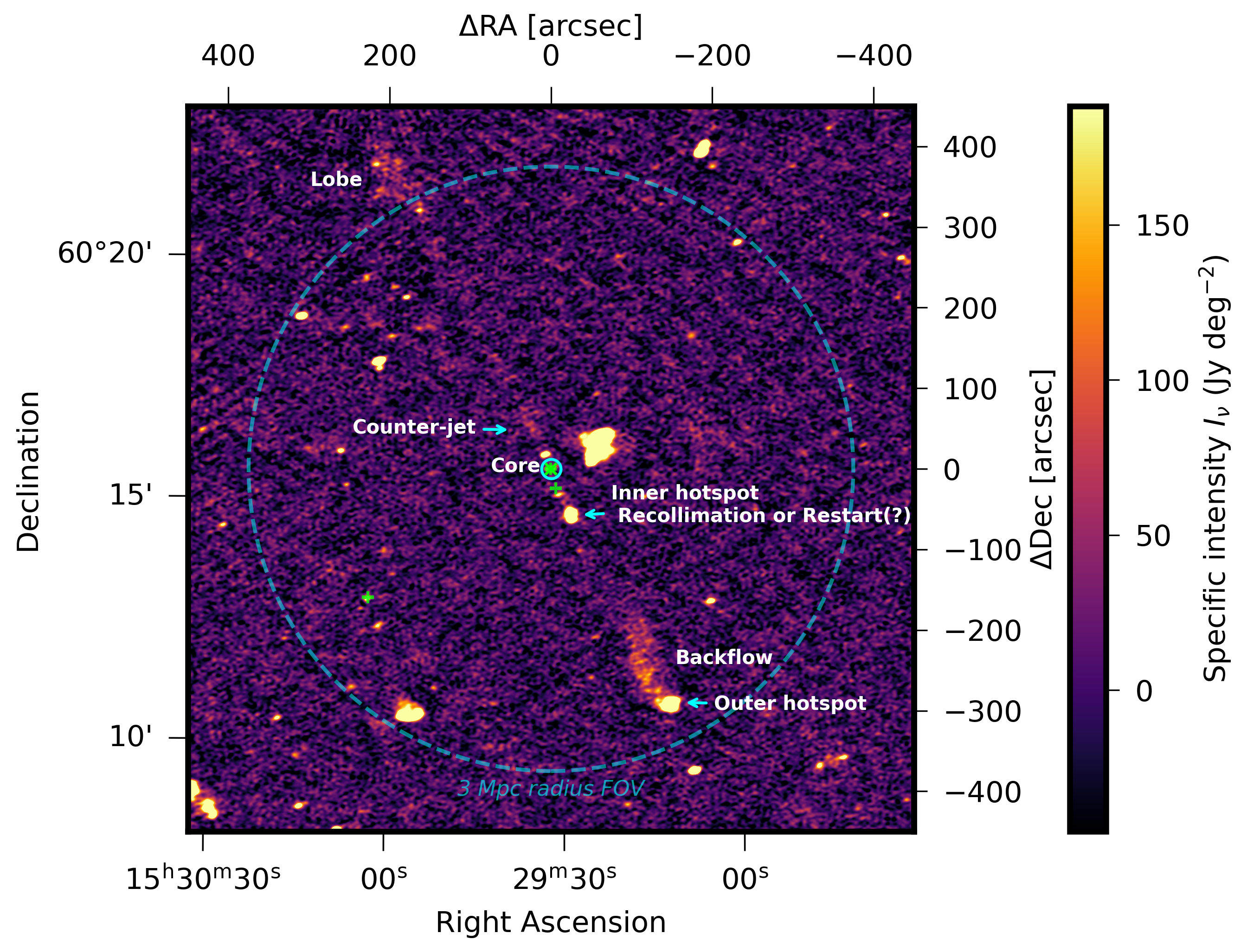}
\caption{J152932.16+601534.4, reported as the largest known extragalactic jet propagation to date (projected length $\sim 6.43$ Mpc; $z_{spec} = 0.896$), exhibiting a high axial-ratio structure (thin cocoon morphology). The observed morphology closely resembles our simulated ``Case A'' (Fig.~\ref{Fig:8Mpc_collage}), including enhanced emission at the jet base and at the jet head. Green plus signs indicate nearby galaxies within a 3 Mpc radius field of view (FOV) and under the less stringent redshift constraint, $\delta z < 0.05$, indicating the source resides in a poor environment.}
\label{Fig:Porphyron}
\end{figure*}

\begin{figure*}[h]
\centering
\includegraphics[width=0.75\textwidth]{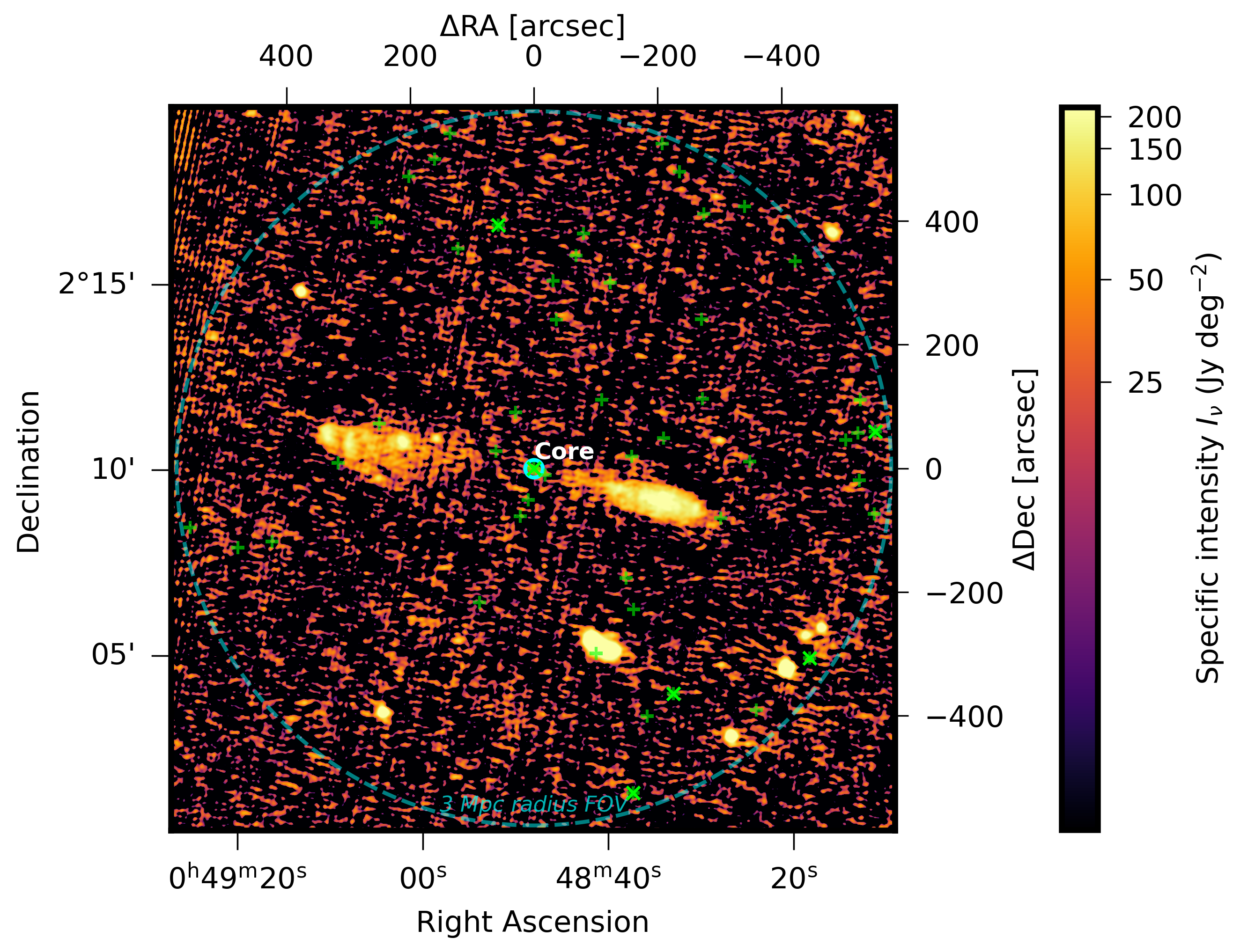}\\
\includegraphics[width=0.75\textwidth]{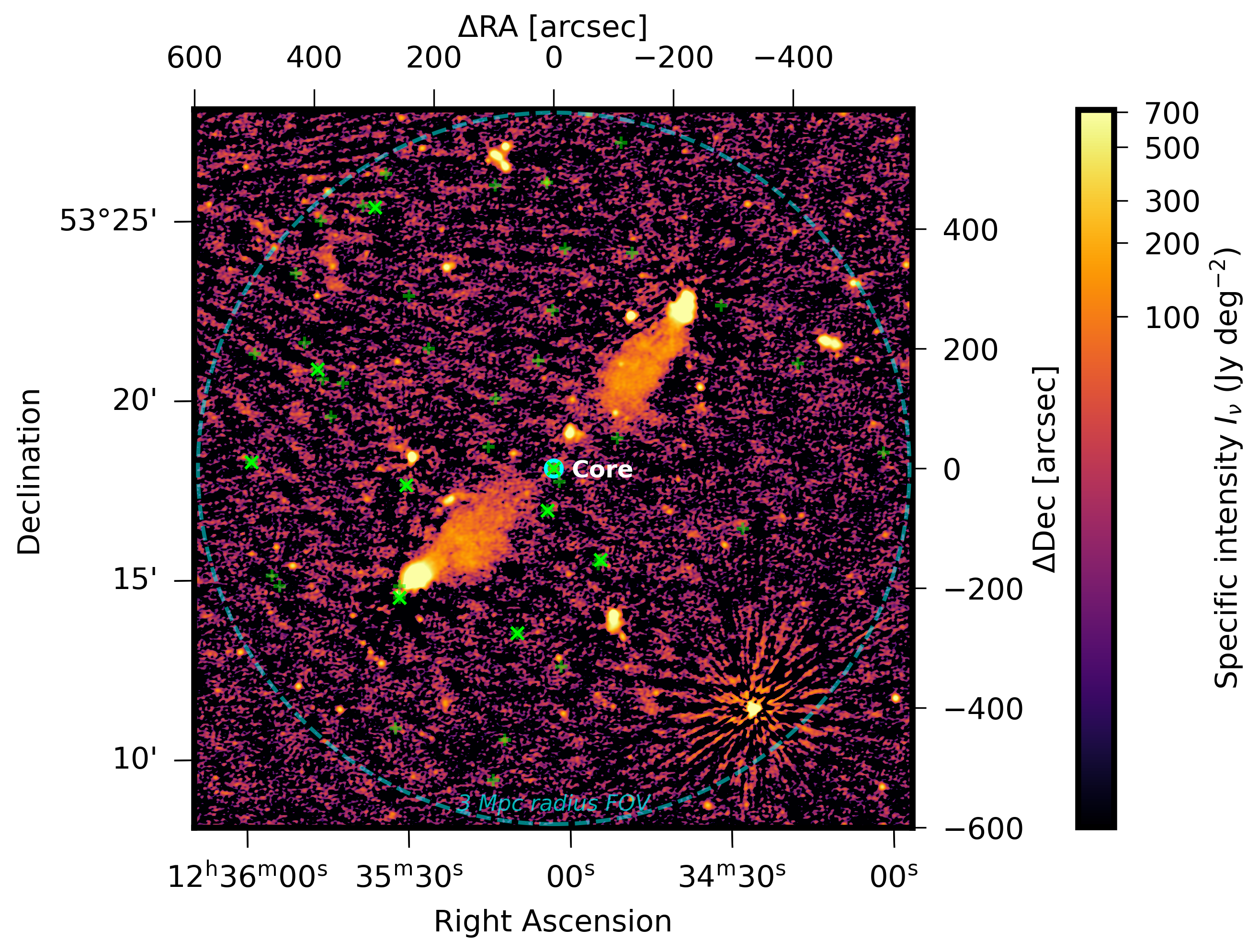}
\caption{J0048+0210 (\textit{top}) and J1235+5317 (\textit{bottom}) are two GRGs with projected sizes of 3.41 Mpc ($z_{spec} = 0.3605$) and 3.25 Mpc ($z_{spec} = 0.3448$), respectively (144 MHz LOFAR maps). Both sources evolve within galaxy clusters where the host galaxy reported to be the brightest cluster member, showing (morphological) similarities to our ``Case B'' and ``C'' simulation scenarios (Fig.~\ref{Fig:Collage_Typical_Cluster}, ~\ref{Fig:Cluster_collage}). The green `×' and `+' markers indicate the surrounding galaxy distribution with our adopted redshift error shells relative to the host ($\delta z < 0.025$, and $\delta z < 0.05$, respectively). All satisfy our selection criteria used to identify relevant environmental overdensity; circles indicate 3 Mpc radius FOVs.}
\label{Fig:GRG-BCGs}
\end{figure*}

We searched for GRGs showing morphologies resembling those of our numerical simulations in the LOFAR Two-metre Sky Survey Third Data Release (LoTSS-DR3 \citep{Shimwell2026}). In addition to Porphyron (J152932.16+601534.4), whose radio image is shown in Fig.~\ref{Fig:Porphyron} (data obtained and replotted with permission from \citet{Oei2024_7Mpc}), we retrieved radio images of two additional sources (from the DR-3 database), namely J0048+0210 and J1235+531 (reported in \citet{Andernach2025}). These are illustrated in Fig.~\ref{Fig:GRG-BCGs}. Host-galaxy spectroscopic redshifts are available for all three sources. We also queried the photometric redshift catalog by \citet{Duncan2022} to investigate the environments in which these observed GRGs reside. We only retained sources in the catalog that satisfied the following criteria:

\begin{itemize}
 \item source location within a projected circle of 3 Mpc radius centered on the host of the GRG \citep[e.g.,][]{Malarecki2015}
 \item source redshift ($z$) consistent within $1\sigma$ of that of the host ($z_{\rm host}$), i.e., $|z - z_{\rm host}| / \sqrt{\delta z^2 + \delta z_{\rm host}^2} < 1.0$
 \item the error of the source's redshift estimate is smaller than a given threshold, that is, $\delta z < \delta z_{\rm max}$
 \item low likelihood of being a star, that is $P_{\rm star} < 0.2$ \citep{Duncan2022}
\end{itemize}

We adopted two redshift matching criteria: a stringent threshold of $\delta z_{\rm max}=0.025$ to identify high-confidence counterparts (green cross symbols), and a more permissive threshold of $\delta z_{\rm max}=0.05$ to include plausible associated sources (faded green plus symbols).
The choice to adopt $\delta z$ (the third constraint noted above) was made to exclude galaxies that match the other criteria in the list, but have large uncertainties in the photometric redshift estimates. In this way, galaxies more likely to lie at redshifts similar to those of the GRG are selected. The locations 
of the accepted associated galaxies are shown in Figs.~\ref{Fig:Porphyron} and \ref{Fig:GRG-BCGs}.

Porphyrion, with its extraordinary projected extent, lies in an exceptionally poor environment, with only two nearby member galaxies identified in its vicinity (Fig.~\ref{Fig:Porphyron}). This observational result is consistent with  
``Case~A'' (Fig.~\ref{Fig:8Mpc_collage}), where the low ambient density permits sustained rapid jet propagation to extreme scales. In contrast, the other two systems shown in Fig.~\ref{Fig:GRG-BCGs}, whose pairs of jets appear to stall at scales of $\sim3$~Mpc, closely respectively resemble ``Cases~B'' and ``C'' in terms of morphology, inferred environmental density, and overall source sizes (Fig.~\ref{Fig:Collage_Typical_Cluster},~\ref{Fig:Cluster_collage}). 

These comparisons support our key conclusions that, while multi-Mpc radio galaxies can form in relatively rich environments through prolonged episodes of jet activity, extreme GRGs exceeding $\sim5$~Mpc likely require the exceptionally rarefied conditions found in the most underdense regions of the cosmic web. Such sources therefore provide a unique opportunity to probe a largely unexplored regime of the low-density Universe.

\subsection{Understanding the scarcity of such extreme radio galaxies}

We have demonstrated that the propagation of extragalactic relativistic jet pairs out to sizes $\gtrsim 7$ Mpc is achievable for reasonable jet and ambient medium parameters. 
However, observational detection of such systems is rare: only a few sources extending beyond 5 Mpc are known \citep{Andernach2025}. While around a hundred sources exceeding 3 Mpc are known they are a small fraction of the total population of GRGs \citep{Mostert2024}. This apparent discrepancy between theoretical feasibility and observational scarcity may be understood through a combination of physical and observational constraints.

From a dynamical perspective, only sufficiently powerful jets are capable of sustaining long-range propagation, as lower-power jets are increasingly susceptible to disruption through interactions with the ambient medium across multiple scales, from the galactic medium to group and cluster environments \citep[e.g.,][]{Mukherjee2016,Giri2026_5Mpc}. The simultaneous requirements of a very high 
kinetic jet power being launched into an underdense void environment capable of supporting jet head propagation $\sim 0.72c$ (our `Case A') certainly occupies a small volume in the phase-space of extragalactic jet propagation parameters \citep[e.g.,][]{Jamrozy2008,An2012,Walg2020}, shedding light on the scarcity of extreme GRGs. This is consistent with studies of the radio luminosity function, which show a steep decline in source number density with increasing jet power \citep[e.g.][]{Ocran2021}. 

At the same time, radiative processes play a critical role: adiabatic expansion losses, synchrotron cooling, as well as inverse Compton scattering off cosmic microwave background photons, significantly reduce the observable emission from extended sources, thereby contributing to their non-detection. This interpretation is supported by both observational and numerical studies of GRGs \citep[e.g.][]{Schoenmakers2000,Giri2025_emission}. Notably, the number of known $> 3$ Mpc sources has increased substantially---from only a few reported prior to 2021 to over a hundred today---primarily driven by the improved sensitivity of modern radio facilities, highlighting the strong role of observational sensitivities \citep{Koribalski2026}. 

It is worth noting that extreme GRGs
may still be produced through prolonged, uninterrupted jet activity in denser media.  But our results for ``Cases~B'' and ``C'' suggest that achieving such sizes in such environments either requires jet powers significantly exceeding those adopted in this study or demands sustained jet injection over much longer timescales. The latter scenario relies on the availability of a stable and efficient fueling mechanism capable of maintaining accretion onto the central engine for $\gtrsim 100$ Myr, which is not generally expected \citep{Lalakos2022,Cammelli2026,Dabhade2025,Elley2026}. So while lengthy continuous jet activity remains a viable pathway for the formation of extreme GRGs, it likely represents a relatively rare combination of favorable energetic and fueling conditions.

It is also important to note that denser cosmic environments are continuously evolving and unlikely to remain perfectly relaxed over extended periods. Such environments are expected to be impacted by a variety of dynamical processes, including turbulent motions, bulk flows, vortices, and winds \citep[cf.,][]{Burns2002}. Indeed, recent observations with XRISM have begun to reveal direct evidence for multi-scaled gas motions \citep{Xrism2025,Heinrich2026,Vazza2026}. These forms of ``cosmic weather" can significantly impede the free propagation of jets, further limiting their ability to attain extreme sizes \citep{Dominguez2024,Cotton2025}. While the influence of such environmental disturbances is beyond the scope of the present work and will be investigated in a dedicated future study, they provide an additional explanation for the rarity of extreme GRGs.

Furthermore, the suppression of MHD instabilities appears to be a crucial requirement for sustaining jet propagation over several megaparsecs. As demonstrated by our simulations \citep[also,][]{Giri2026_5Mpc}, the onset of jet decollimation and growth of instabilities  can substantially reduce the advance speed and overall extent reached by these extreme 
sources. Since the development of such instabilities is highly likely in extended RGs \citep{Guan2014,Tchekhovskoy2016,Perucho2019,Wang2023,Costa2026}, the simultaneous fulfillment of both favorable environmental conditions and long-term jet stability is certainly uncommon, contributing further to the observed rarity of extreme GRGs. To elucidate how high powered jets suppress these destructive lateral undulations along the beam, we presented a spectral stability analysis in Section~\ref{Sec:Jet inertia against lateral instabilities} (and Fig.~\ref{Fig:FFT_LorG}).

A final consideration is the role of projection effects \citep{Hodges-kluck2011}, which can significantly underestimate the true linear extent of radio galaxies. For example, the reported GRQ with a projected size of 4.45~Mpc would have an intrinsic extent of approximately 6.3~Mpc for an inclination angle of $45^\circ$ to the line of sight and could be substantially larger \citep{Coziol2017}. Likewise, Prophyrion, with a projected size of 6.43~Mpc, is inferred from deprojection analyses to possess an intrinsic extent exceeding 7~Mpc \citep{Oei2024_7Mpc}. Many GRGs exhibit pronounced asymmetries in arm length, lobe brightness, and overall morphology \citep{Schoenmakers2000}, even in restarted sources displaying evidence of asymmetric renewed jet components \citep{Dabhade2025}. These characteristics hint that a substantial fraction of GRGs are not oriented close to the plane of the sky \citep[see also,][]{Walg2020,Oei2023}, implying that their observed sizes represent lower limits to their intrinsic extents. 
Projection effects are certainly important to consider when assessing the apparent rarity of the most extreme GRGs, as the number of intrinsically larger systems is observationally underestimated.

Within this context, our modelling represents a first systematic exploration of a previously under-sampled and computationally challenging parameter space, enabled by modern numerical techniques and powerful supercomputers. The results provide a proof-of-concept that such large-scale jet propagation is physically plausible within standard jet frameworks, thereby reinforcing conventional models and reducing the need to invoke more exotic explanations \citep[e.g.,][]{Neronov2025}.

Nevertheless, a comprehensive understanding of the rarity of these systems requires further investigation. This should include a broader exploration of the jet–environment parameter space, such as varying jet-to-ambient density contrasts, flow velocities, injection variability, and the resolution of MHD instabilities. It should eventually also incorporate second-order environmental effects. These include interactions with multiphase media, dense clumps, turbulent motions, and gradients across multi-scale environments, all of which can enhance entrainment and limit jet propagation. Capturing these processes pushes simulations into an unprecedented computational regime, posing significant challenges even for GPU-accelerated frameworks. As such, a dedicated and carefully designed simulation campaign is essential to more fully elucidate 
the physical origins of the observed scarcity of these extreme GRGs.

\section{Summary}\label{Sec:Summary}
Jetted radio galaxies with spatial extents typically exceeding $1\text{ Mpc}$ are classified as giant radio galaxies, still representing a rare subclass of the broader radio galaxy population. Their vast propagation scales make GRGs ideal probes for studying cosmic-scale energy transport, acting as giant barometers that map ambient pressure profiles across cosmic overdensities. 
The recent discovery of ``extreme'' GRGs spanning over $5\text{ Mpc}$, however, has pushed the frontiers of extragalactic jet research—making them a key topic of contemporary interest despite the handful number of known systems. These extreme systems challenge conventional understandings by raising critical questions about prolonged (gigayear-scale) accretion event at the black hole scales, anomalously high propagation velocities that suppress fluid instabilities, and jet evolution within the lowest-density intergalactic medium, to name a few.
The recent discovery of Porphyrion (extending nearly $7\text{ Mpc}$) has revived these questions, prompting proposed alternative origins such as gamma-ray beam (originating from host AGN) dissipating within intergalactic magnetic fields. 

In this work, we examine the propagation dynamics and dissipation scales of relativistic jets to identify the physical conditions that foster extreme giant radio galaxies and map the parameter space for even larger jetted sources awaiting discovery. To achieve this, we employ a computational domain significantly larger than previous numerical studies in this regime. This work thus represents the first high-resolution (zoom-in) simulation tracking jet evolution to cosmic-web scales, made feasible by leveraging GPU-accelerated software on world-class high-performance computing infrastructure.

Our findings reveal that when reasonable jet power and low-density environmental conditions act in tandem, the upper limit on extragalactic jet propagation scales remains unconstrained. This implies that the currently observed maximum source dimensions do not represent a strict physical ceiling, predicting an undiscovered population of even larger giant radio galaxies. The apparent rarity of such extreme sources is further accounted for by delineating key physical and observational factors that limit their generation, appearance and detection. Nevertheless, these distinct megaparsec-scale jets represent physically viable systems rather than statistical anomalies. 

We justify our suite of numerical runs—which isolate environmental variations at a constant jet power, complementing our earlier work on jet power scaling—by comparing them with observed GRG properties across relevant scales. Using these observationally benchmarked setups, we examine the jet spine dynamics, uncovering critical mesoscale thermodynamic and magnetohydrodynamic mechanisms that regulate the propagation and radiative appearance of the largest galactic-driven outflows in the Universe.

Although this work provides an unprecedented high-resolution view of jet propagation and dissipation on cosmic scales, several physical mechanisms warrant further exploration. In particular, incorporating the secondary drivers of jet disruption—such as accretion-driven spatio-temporal engine variability, clumpy and unrelaxed environments, intricate jet-injection dynamics, alongside, sub-grid radiative processes controlling the emission fingerprints, and alternative models like restarted jet activity—will require dedicated future supercomputing campaigns to push beyond current computational boundaries, which will be the focus of our upcoming simulation campaigns.

\begin{acknowledgments}
We acknowledge the Istituto Nazionale di AstroFisica (INAF) for awarding this project access and computational time to the LEONARDO supercomputer, owned by the EuroHPC Joint Undertaking, hosted by CINECA (Italy) and the LEONARDO consortium.
Scientific results presented in this work were obtained with the aid of the gPLUTO code, for which the authors acknowledge funding from the European High Performance Computing Joint Undertaking (JU) and Belgium, Czech Republic, France, Germany, Greece, Italy, Norway, and Spain under grant agreement No. 101093441 (SPACE). The authors thank the SPACE consortium for making the simulation code publicly available. LOFAR is the Low Frequency Array designed and constructed by ASTRON. It has observing, data processing, and data storage facilities in several countries, which are owned by various parties (each with their own funding sources), and which are collectively operated by the LOFAR ERIC under a joint scientific policy. The LOFAR resources have benefited from the following recent major funding sources: CNRS-INSU, Observatoire de Paris and Université d'Orléans, France; BMFTR, MKW-NRW, MPG, Germany; Science Foundation Ireland (SFI), Department of Business, Enterprise and Innovation (DBEI), Ireland; NWO, The Netherlands; The Science and Technology Facilities Council, UK; Ministry of Science and Higher Education, Poland; The Istituto Nazionale di Astrofisica (INAF), Italy.
Part of this work has been carried out within the framework of the EUROfusion Consortium, funded by the European Union via the Euratom Research and Training Programme (Grant Agreement No. 101052200 – EUROfusion). Computational resources were provided through the EUROfusion High-Performance Computing (HPC) Programme (Project FUPB1 T-RECS, EUROfusion PITAGORA BOOSTER Call 9) on the Pitagora high-performance computing facility at CINECA, Italy.
D.B. gratefully acknowledges Dr. D. Molinari (CINECA) for his valuable assistance in optimizing our use of the PITAGORA HPC system reservation, and Prof. M. Muraglia, Principal Investigator of the T-RECS project, for granting access to the T-RECS computational allocation on the PITAGORA system.
\end{acknowledgments}


\bibliography{sample701}{}

\begin{thebibliography}{}
\expandafter\ifx\csname natexlab\endcsname\relax\def\natexlab#1{#1}\fi
\providecommand{\url}[1]{\href{#1}{#1}}
\providecommand{\dodoi}[1]{doi:~\href{http://doi.org/#1}{\nolinkurl{#1}}}
\providecommand{\doeprint}[1]{\href{http://ascl.net/#1}{\nolinkurl{http://ascl.net/#1}}}
\providecommand{\doarXiv}[1]{\href{https://arxiv.org/abs/#1}{\nolinkurl{https://arxiv.org/abs/#1}}}

\bibitem[{T. {An} {et~al.}(2012){An}, {Wu}, {Yang}, {Taylor}, {Hong}, {Baan}, {Liu}, {Wang}, {Zhang}, {Wang}, {Chen}, {Cui}, {Hao}, \& {Zhu}}]{An2012}
{An}, T., {Wu}, F., {Yang}, J., {et~al.} 2012, \bibinfo{title}{{VLBI Observations of 10 Compact Symmetric Object Candidates: Expansion Velocities of Hot Spots},} \apjs, 198, 5, \dodoi{10.1088/0067-0049/198/1/5}

\bibitem[{H. {Andernach} \& M. {Br{\"u}ggen}(2025){Andernach} \& {Br{\"u}ggen}}]{Andernach2025}
{Andernach}, H., \& {Br{\"u}ggen}, M. 2025, \bibinfo{title}{{Properties of giant radio galaxies larger than 3 Mpc},} \aap, 699, A257, \dodoi{10.1051/0004-6361/202452961}

\bibitem[{P.~D. {Barthel}(1989){Barthel}}]{Barthel1989}
{Barthel}, P.~D. 1989, \bibinfo{title}{{Is Every Quasar Beamed?},} \apj, 336, 606, \dodoi{10.1086/167038}

\bibitem[{J.~O. {Burns} {et~al.}(2002){Burns}, {Loken}, {Roettiger}, {Rizza}, {Bryan}, {Norman}, {G{\'o}mez}, \& {Owen}}]{Burns2002}
{Burns}, J.~O., {Loken}, C., {Roettiger}, K., {et~al.} 2002, \bibinfo{title}{{Stormy weather and cluster radio galaxies},} \nar, 46, 135, \dodoi{10.1016/S1387-6473(01)00168-3}

\bibitem[{V. {Cammelli} {et~al.}(2026){Cammelli}, {Gaspari}, {Barbani}, {Piana}, {Stel}, {Brustio}, {Olivares}, {Serafinelli}, {Temi}, {Salvestrini}, {Reefe}, {Maccagni}, \& {Tombesi}}]{Cammelli2026}
{Cammelli}, V., {Gaspari}, M., {Barbani}, F., {et~al.} 2026, \bibinfo{title}{{BlackHoleWeather -- Jet-regulated chaotic cold accretion across the meso scale: Variability and kinematics},} arXiv e-prints, arXiv:2605.27511.
\newblock \doarXiv{2605.27511}

\bibitem[{T.~M. {Cantwell} {et~al.}(2020){Cantwell}, {Bray}, {Croston}, {Scaife}, {Mulcahy}, {Best}, {Br{\"u}ggen}, {Brunetti}, {Callingham}, {Clarke}, {Hardcastle}, {Harwood}, {Heald}, {Heesen}, {Iacobelli}, {Jamrozy}, {Morganti}, {Orr{\'u}}, {O'Sullivan}, {Riseley}, {R{\"o}ttgering}, {Shulevski}, {Sridhar}, {Tasse}, \& {Van Eck}}]{Cantwell2020}
{Cantwell}, T.~M., {Bray}, J.~D., {Croston}, J.~H., {et~al.} 2020, \bibinfo{title}{{Low-frequency observations of the giant radio galaxy NGC 6251},} \mnras, 495, 143, \dodoi{10.1093/mnras/staa1160}

\bibitem[{K.~W. {Cavagnolo} {et~al.}(2008){Cavagnolo}, {Donahue}, {Voit}, \& {Sun}}]{Cavagnolo2008}
{Cavagnolo}, K.~W., {Donahue}, M., {Voit}, G.~M., \& {Sun}, M. 2008, \bibinfo{title}{{An Entropy Threshold for Strong H{\ensuremath{\alpha}} and Radio Emission in the Cores of Galaxy Clusters},} \apjl, 683, L107, \dodoi{10.1086/591665}

\bibitem[{A. {Cavaliere} \& R. {Fusco-Femiano}(1976){Cavaliere} \& {Fusco-Femiano}}]{Cavaliere1976}
{Cavaliere}, A., \& {Fusco-Femiano}, R. 1976, \bibinfo{title}{{X-rays from hot plasma in clusters of galaxies.},} \aap, 49, 137

\bibitem[{K. {Chavan} {et~al.}(2023){Chavan}, {Dabhade}, \& {Saikia}}]{Chavan2023}
{Chavan}, K., {Dabhade}, P., \& {Saikia}, D.~J. 2023, \bibinfo{title}{{A giant radio galaxy with three cycles of episodic jet activity from LoTSS DR2},} \mnras, 525, L87, \dodoi{10.1093/mnrasl/slad100}

\bibitem[{Y. {Chen} {et~al.}(2007){Chen}, {Reiprich}, {B{\"o}hringer}, {Ikebe}, \& {Zhang}}]{Chen2007}
{Chen}, Y., {Reiprich}, T.~H., {B{\"o}hringer}, H., {Ikebe}, Y., \& {Zhang}, Y.-Y. 2007, \bibinfo{title}{{Statistics of X-ray observables for the cooling-core and non-cooling core galaxy clusters},} \aap, 466, 805, \dodoi{10.1051/0004-6361:20066471}

\bibitem[{S. {Cielo} {et~al.}(2017){Cielo}, {Antonuccio-Delogu}, {Silk}, \& {Romeo}}]{Cielo2017}
{Cielo}, S., {Antonuccio-Delogu}, V., {Silk}, J., \& {Romeo}, A.~D. 2017, \bibinfo{title}{{Backflows by active galactic nuclei jets: global properties and influence on supermassive black hole accretion},} \mnras, 467, 4526, \dodoi{10.1093/mnras/stx403}

\bibitem[{A. {Costa} {et~al.}(2026){Costa}, {Bodo}, {Tavecchio}, {Rossi}, {Coppi}, {Sciaccaluga}, \& {Boula}}]{Costa2026}
{Costa}, A., {Bodo}, G., {Tavecchio}, F., {et~al.} 2026, \bibinfo{title}{{How do recollimation-induced instabilities shape the propagation of hydrodynamic relativistic jets?},} \aap, 705, A74, \dodoi{10.1051/0004-6361/202554698}

\bibitem[{W.~D. {Cotton} {et~al.}(2025){Cotton}, {Giri}, {Agnihotri}, {Saikia}, {Thorat}, \& {Camilo}}]{Cotton2025}
{Cotton}, W.~D., {Giri}, G., {Agnihotri}, P.~J., {et~al.} 2025, \bibinfo{title}{{MeerKAT 1.3 GHz Observations of the Wide-angle Tail Radio Galaxy J1712-2435},} \apj, 989, 223, \dodoi{10.3847/1538-4357/adf186}

\bibitem[{R. {Coziol} {et~al.}(2017){Coziol}, {Andernach}, {Torres-Papaqui}, {Ortega-Minakata}, \& {Moreno del Rio}}]{Coziol2017}
{Coziol}, R., {Andernach}, H., {Torres-Papaqui}, J.~P., {Ortega-Minakata}, R.~A., \& {Moreno del Rio}, F. 2017, \bibinfo{title}{{What sparks the radio-loud phase of nearby quasars?},} \mnras, 466, 921, \dodoi{10.1093/mnras/stw3164}

\bibitem[{P. {Dabhade} {et~al.}(2025){Dabhade}, {Chavan}, {Saikia}, {Oei}, \& {R{\"o}ttgering}}]{Dabhade2025}
{Dabhade}, P., {Chavan}, K., {Saikia}, D.~J., {Oei}, M.~S.~S.~L., \& {R{\"o}ttgering}, H.~J.~A. 2025, \bibinfo{title}{{Search and analysis of giant radio galaxies with associated nuclei (SAGAN): V. Study of giant double-double radio galaxies from LoTSS DR2},} \aap, 696, A97, \dodoi{10.1051/0004-6361/202451918}

\bibitem[{P. {Dabhade} {et~al.}(2023){Dabhade}, {Saikia}, \& {Mahato}}]{Dabhade2023}
{Dabhade}, P., {Saikia}, D.~J., \& {Mahato}, M. 2023, \bibinfo{title}{{Decoding the giant extragalactic radio sources},} Journal of Astrophysics and Astronomy, 44, 13, \dodoi{10.1007/s12036-022-09898-5}

\bibitem[{P. {Dabhade} {et~al.}(2020){Dabhade}, {Mahato}, {Bagchi}, {Saikia}, {Combes}, {Sankhyayan}, {R{\"o}ttgering}, {Ho}, {Gaikwad}, {Raychaudhury}, {Vaidya}, \& {Guiderdoni}}]{Dabhade2020_SAGAN}
{Dabhade}, P., {Mahato}, M., {Bagchi}, J., {et~al.} 2020, \bibinfo{title}{{Search and analysis of giant radio galaxies with associated nuclei (SAGAN). I. New sample and multi-wavelength studies},} \aap, 642, A153, \dodoi{10.1051/0004-6361/202038344}

\bibitem[{R.~A. {Daly} {et~al.}(2012){Daly}, {Sprinkle}, {O'Dea}, {Kharb}, \& {Baum}}]{Daly2012}
{Daly}, R.~A., {Sprinkle}, T.~B., {O'Dea}, C.~P., {Kharb}, P., \& {Baum}, S.~A. 2012, \bibinfo{title}{{The relationship between beam power and radio power for classical double radio sources},} \mnras, 423, 2498, \dodoi{10.1111/j.1365-2966.2012.21060.x}

\bibitem[{A. {Dedner} {et~al.}(2002){Dedner}, {Kemm}, {Kr{\"o}ner}, {Munz}, {Schnitzer}, \& {Wesenberg}}]{Dedner2002}
{Dedner}, A., {Kemm}, F., {Kr{\"o}ner}, D., {et~al.} 2002, \bibinfo{title}{{Hyperbolic Divergence Cleaning for the MHD Equations},} Journal of Computational Physics, 175, 645, \dodoi{10.1006/jcph.2001.6961}

\bibitem[{P. {Dom{\'\i}nguez-Fern{\'a}ndez} {et~al.}(2024){Dom{\'\i}nguez-Fern{\'a}ndez}, {ZuHone}, {Weinberger}, {Bellomi}, {Hernquist}, {Nulsen}, \& {Brunetti}}]{Dominguez2024}
{Dom{\'\i}nguez-Fern{\'a}ndez}, P., {ZuHone}, J., {Weinberger}, R., {et~al.} 2024, \bibinfo{title}{{Jet Interaction with Galaxy Cluster Mergers},} \apj, 977, 221, \dodoi{10.3847/1538-4357/ad9028}

\bibitem[{R.~P. {Dubey} {et~al.}(2023){Dubey}, {Fendt}, \& {Vaidya}}]{Dubey2023}
{Dubey}, R.~P., {Fendt}, C., \& {Vaidya}, B. 2023, \bibinfo{title}{{Particles in Relativistic MHD Jets. I. Role of Jet Dynamics in Particle Acceleration},} \apj, 952, 1, \dodoi{10.3847/1538-4357/ace0bf}

\bibitem[{R.~P. {Dubey} {et~al.}(2024){Dubey}, {Fendt}, \& {Vaidya}}]{Dubey2024}
{Dubey}, R.~P., {Fendt}, C., \& {Vaidya}, B. 2024, \bibinfo{title}{{Particles in Relativistic Magnetohydrodynamic Jets. II. Bridging Jet Dynamics with Multi-wave band Nonthermal Emission Signatures},} \apj, 976, 144, \dodoi{10.3847/1538-4357/ad8135}

\bibitem[{K.~J. {Duncan}(2022){Duncan}}]{Duncan2022}
{Duncan}, K.~J. 2022, \bibinfo{title}{{All-purpose, all-sky photometric redshifts for the Legacy Imaging Surveys Data Release 8},} \mnras, 512, 3662, \dodoi{10.1093/mnras/stac608}

\bibitem[{E.~L. {Elley} {et~al.}(2026){Elley}, {Matthews}, {Whitehead}, \& {Cooper}}]{Elley2026}
{Elley}, E.~L., {Matthews}, J.~H., {Whitehead}, H., \& {Cooper}, A.~J. 2026, \bibinfo{title}{{The impact of flickering variability and magnetisation on the dynamics, stability and morphology of radio-loud AGN jets},} arXiv e-prints, arXiv:2605.13469, \dodoi{10.48550/arXiv.2605.13469}

\bibitem[{B.~L. {Fanaroff} \& J.~M. {Riley}(1974){Fanaroff} \& {Riley}}]{Fanaroff1974}
{Fanaroff}, B.~L., \& {Riley}, J.~M. 1974, \bibinfo{title}{{The morphology of extragalactic radio sources of high and low luminosity},} \mnras, 167, 31P, \dodoi{10.1093/mnras/167.1.31P}

\bibitem[{M. {Gaspari} {et~al.}(2020){Gaspari}, {Tombesi}, \& {Cappi}}]{Gaspari2020}
{Gaspari}, M., {Tombesi}, F., \& {Cappi}, M. 2020, \bibinfo{title}{{Linking macro-, meso- and microscales in multiphase AGN feeding and feedback},} Nature Astronomy, 4, 10, \dodoi{10.1038/s41550-019-0970-1}

\bibitem[{S. {Giacintucci} {et~al.}(2011){Giacintucci}, {Dallacasa}, {Venturi}, {Brunetti}, {Cassano}, {Markevitch}, \& {Athreya}}]{Giacintucci2011}
{Giacintucci}, S., {Dallacasa}, D., {Venturi}, T., {et~al.} 2011, \bibinfo{title}{{An unlikely radio halo in the low X-ray luminosity galaxy cluster RXC J1514.9-1523},} \aap, 534, A57, \dodoi{10.1051/0004-6361/201117820}

\bibitem[{G. {Giovannini} {et~al.}(2001){Giovannini}, {Cotton}, {Feretti}, {Lara}, \& {Venturi}}]{Giovannini2001}
{Giovannini}, G., {Cotton}, W.~D., {Feretti}, L., {Lara}, L., \& {Venturi}, T. 2001, \bibinfo{title}{{VLBI Observations of a Complete Sample of Radio Galaxies: 10 Years Later},} \apj, 552, 508, \dodoi{10.1086/320581}

\bibitem[{G. {Giovannini} {et~al.}(1999){Giovannini}, {Taylor}, {Arbizzani}, {Bondi}, {Cotton}, {Feretti}, {Lara}, \& {Venturi}}]{Giovannini1999}
{Giovannini}, G., {Taylor}, G.~B., {Arbizzani}, E., {et~al.} 1999, \bibinfo{title}{{B2 1144+35: A Giant Low-Power Radio Galaxy with Superluminal Motion},} \apj, 522, 101, \dodoi{10.1086/307640}

\bibitem[{G. {Giri} {et~al.}(2025{\natexlab{a}}){Giri}, {Bagchi}, {Thorat}, {Deane}, {Delhaize}, \& {Saikia}}]{Giri2025_dynamics}
{Giri}, G., {Bagchi}, J., {Thorat}, K., {et~al.} 2025{\natexlab{a}}, \bibinfo{title}{{Probing the formation of megaparsec-scale giant radio galaxies: I. Dynamical insights from magnetohydrodynamic simulations},} \aap, 693, A77, \dodoi{10.1051/0004-6361/202451812}

\bibitem[{G. {Giri} {et~al.}(2025{\natexlab{b}}){Giri}, {Fendt}, {Bagchi}, {Thorat}, {Saikia}, {Deane}, \& {Delhaize}}]{Giri2025_emission}
{Giri}, G., {Fendt}, C., {Bagchi}, J., {et~al.} 2025{\natexlab{b}}, \bibinfo{title}{{Probing the formation of megaparsec-scale giant radio galaxies: II. Continuum and polarization behavior from magneto-hydrodynamic simulations},} \aap, 703, A214, \dodoi{10.1051/0004-6361/202554872}

\bibitem[{G. {Giri} {et~al.}(2026{\natexlab{a}}){Giri}, {Mignone}, {Suriano}, {Rossazza}, \& {Truzzi}}]{Giri2026}
{Giri}, G., {Mignone}, A., {Suriano}, A., {Rossazza}, M., \& {Truzzi}, S. 2026{\natexlab{a}}, \bibinfo{title}{{Performance assessment of the gPLUTOcode for the numerical modeling of radio galaxy evolution},} Astronomy and Computing, 56, 101091, \dodoi{10.1016/j.ascom.2026.101091}

\bibitem[{G. {Giri} {et~al.}(2026{\natexlab{b}}){Giri}, {Ghosh}, {Joshi}, {Caproni}, {Rossi}, {Bodo}, {Kundu}, {Thorat}, {Chatterjee}, {Borgogno}, {Vittorini}, \& {Tavani}}]{Giri2026_RGs}
{Giri}, G., {Ghosh}, P., {Joshi}, R., {et~al.} 2026{\natexlab{b}}, \bibinfo{title}{{Galactic-scale evolution of classical and complex radio galaxies: Impact of ambient morphology and jet geometry},} \aap, 709, A279, \dodoi{10.1051/0004-6361/202658848}

\bibitem[{G. {Giri} {et~al.}(2026{\natexlab{c}}){Giri}, {Borgogno}, {Tavani}, {Mignone}, {Sharma}, {Gheller}, {Vittorini}, {Wiita}, {Fanaroff}, {Suriano}, {Saikia}, \& {Brunetti}}]{Giri2026_5Mpc}
{Giri}, G., {Borgogno}, D., {Tavani}, M., {et~al.} 2026{\natexlab{c}}, \bibinfo{title}{{Simulating megaparsec-scale jets of radio galaxies: magnetohydrodynamics of jets reaching 5 Mpc},} \mnras, 550, stag1243, \dodoi{10.1093/mnras/stag1243}

\bibitem[{X. {Guan} {et~al.}(2014){Guan}, {Li}, \& {Li}}]{Guan2014}
{Guan}, X., {Li}, H., \& {Li}, S. 2014, \bibinfo{title}{{Relativistic MHD Simulations of Poynting Flux-driven Jets},} \apj, 781, 48, \dodoi{10.1088/0004-637X/781/1/48}

\bibitem[{M.~J. {Hardcastle}(2018){Hardcastle}}]{Hardcastle2018}
{Hardcastle}, M.~J. 2018, \bibinfo{title}{{A simulation-based analytic model of radio galaxies},} \mnras, 475, 2768, \dodoi{10.1093/mnras/stx3358}

\bibitem[{M.~J. {Hardcastle} \& M.~G.~H. {Krause}(2013){Hardcastle} \& {Krause}}]{Hardcastle2013}
{Hardcastle}, M.~J., \& {Krause}, M.~G.~H. 2013, \bibinfo{title}{{Numerical modelling of the lobes of radio galaxies in cluster environments},} \mnras, 430, 174, \dodoi{10.1093/mnras/sts564}

\bibitem[{M.~J. {Hardcastle} {et~al.}(2023){Hardcastle}, {Horton}, {Williams}, {Duncan}, {Alegre}, {Barkus}, {Croston}, {Dickinson}, {Osinga}, {R{\"o}ttgering}, {Sabater}, {Shimwell}, {Smith}, {Best}, {Botteon}, {Br{\"u}ggen}, {Drabent}, {de Gasperin}, {G{\"u}rkan}, {Hajduk}, {Hale}, {Hoeft}, {Jamrozy}, {Kunert-Bajraszewska}, {Kondapally}, {Magliocchetti}, {Mahatma}, {Mostert}, {O'Sullivan}, {Pajdosz-{\'S}mierciak}, {Petley}, {Pierce}, {Prandoni}, {Schwarz}, {Shulewski}, {Siewert}, {Stott}, {Tang}, {Vaccari}, {Zheng}, {Bailey}, {Desbled}, {Goyal}, {Gonano}, {Hanset}, {Kurtz}, {Lim}, {Mielle}, {Molloy}, {Roth}, {Terentev}, \& {Torres}}]{Hardcastle2023}
{Hardcastle}, M.~J., {Horton}, M.~A., {Williams}, W.~L., {et~al.} 2023, \bibinfo{title}{{The LOFAR Two-Metre Sky Survey. VI. Optical identifications for the second data release},} \aap, 678, A151, \dodoi{10.1051/0004-6361/202347333}

\bibitem[{M.~J. {Hardcastle} {et~al.}(2026){Hardcastle}, {Konar}, {Dutta}, {Lal}, {Baldi}, {Brienza}, {Hota}, {Kunert-Bajraszewska}, {Mingo}, {Pandey-Pommier}, {Rubinur}, {Shabala}, \& {Vardoulaki}}]{Hardcastle2026}
{Hardcastle}, M.~J., {Konar}, C., {Dutta}, S., {et~al.} 2026, \bibinfo{title}{{Radio Galaxies and Jet Duty Cycles},} arXiv e-prints, arXiv:2606.28755, \dodoi{10.48550/arXiv.2606.28755}

\bibitem[{A. {Heinrich} {et~al.}(2026){Heinrich}, {Zhuravleva}, {Churazov}, {Zhang}, {Rogantini}, {McCall}, {van Weeren}, \& {Forman}}]{Heinrich2026}
{Heinrich}, A., {Zhuravleva}, I., {Churazov}, E., {et~al.} 2026, \bibinfo{title}{{The major cluster merger in Abell 2034 as seen by XRISM: Strong turbulence and spectral anomalies?},} arXiv e-prints, arXiv:2604.27161, \dodoi{10.48550/arXiv.2604.27161}

\bibitem[{E.~J. {Hodges-Kluck} \& C.~S. {Reynolds}(2011){Hodges-Kluck} \& {Reynolds}}]{Hodges-kluck2011}
{Hodges-Kluck}, E.~J., \& {Reynolds}, C.~S. 2011, \bibinfo{title}{{Hydrodynamic Models of Radio Galaxy Morphology: Winged and X-shaped Sources},} \apj, 733, 58, \dodoi{10.1088/0004-637X/733/1/58}

\bibitem[{F. {Hu{\v{s}}ko} {et~al.}(2026){Hu{\v{s}}ko}, {Lacey}, {Schaye}, {Schaller}, {Chaikin}, {Ploeckinger}, {Ben{\'\i}tez Llambay}, {Richings}, \& {Trayford}}]{Husko2026}
{Hu{\v{s}}ko}, F., {Lacey}, C.~G., {Schaye}, J., {et~al.} 2026, \bibinfo{title}{{A hybrid active galactic nucleus feedback model with spinning black holes, winds and jets},} \mnras, 547, stag324, \dodoi{10.1093/mnras/stag324}

\bibitem[{J. {Ineson} {et~al.}(2015){Ineson}, {Croston}, {Hardcastle}, {Kraft}, {Evans}, \& {Jarvis}}]{Ineson2015}
{Ineson}, J., {Croston}, J.~H., {Hardcastle}, M.~J., {et~al.} 2015, \bibinfo{title}{{The link between accretion mode and environment in radio-loud active galaxies},} \mnras, 453, 2682, \dodoi{10.1093/mnras/stv1807}

\bibitem[{M. {Jamrozy} {et~al.}(2008){Jamrozy}, {Konar}, {Machalski}, \& {Saikia}}]{Jamrozy2008}
{Jamrozy}, M., {Konar}, C., {Machalski}, J., \& {Saikia}, D.~J. 2008, \bibinfo{title}{{A multifrequency study of giant radio sources - II. Spectral ageing analysis of the lobes of selected sources},} \mnras, 385, 1286, \dodoi{10.1111/j.1365-2966.2007.12772.x}

\bibitem[{M. {Jamrozy} {et~al.}(2017){Jamrozy}, {Machalski}, {Nikiel-Wroczy{\'n}ski}, \& {Intema}}]{Jamrozy2017}
{Jamrozy}, M., {Machalski}, J., {Nikiel-Wroczy{\'n}ski}, B., \& {Intema}, H.~T. 2017, \bibinfo{title}{{Reassessment of an Origin of the Radio Structure of J1420-0545},} \apj, 850, 7, \dodoi{10.3847/1538-4357/aa9182}

\bibitem[{A.~D. {Kapinska} {et~al.}(2015){Kapinska}, {Hardcastle}, {Jackson}, {An}, {Baan}, \& {Jarvis}}]{Kapinska2015}
{Kapinska}, A.~D., {Hardcastle}, M., {Jackson}, C., {et~al.} 2015, \bibinfo{title}{{Unravelling lifecycles and physics of radio-loud AGN in the SKA Era},} in Advancing Astrophysics with the Square Kilometre Array (AASKA14), 173, \dodoi{10.22323/1.215.0173}

\bibitem[{B.~S. {Koribalski}(2026){Koribalski}}]{Koribalski2026}
{Koribalski}, B.~S. 2026, \bibinfo{title}{{Discoveries of Giant Radio Galaxies demonstrating the power of low surface brightness, wide-field imaging at high resolution},} \pasa, 43, e012, \dodoi{10.1017/pasa.2026.10144}

\bibitem[{S. {Kumari} {et~al.}(2026){Kumari}, {Pal}, {Paul}, \& {Jamrozy}}]{Kumari2026}
{Kumari}, S., {Pal}, S., {Paul}, S., \& {Jamrozy}, M. 2026, \bibinfo{title}{{Probing AGN duty cycle and cluster-driven morphology in a giant episodic radio galaxy},} \mnras, 545, staf2038, \dodoi{10.1093/mnras/staf2038}

\bibitem[{S. {Kundu} {et~al.}(2022){Kundu}, {Vaidya}, {Mignone}, \& {Hardcastle}}]{Kundu2022}
{Kundu}, S., {Vaidya}, B., {Mignone}, A., \& {Hardcastle}, M.~J. 2022, \bibinfo{title}{{A numerical study of the interplay between Fermi acceleration mechanisms in radio lobes of FR-II radio galaxies},} \aap, 667, A138, \dodoi{10.1051/0004-6361/202244251}

\bibitem[{R.~A. {Laing} \& A.~H. {Bridle}(2015){Laing} \& {Bridle}}]{Laing2015}
{Laing}, R.~A., \& {Bridle}, A.~H. 2015, \bibinfo{title}{{Kinematics of kiloparsec-scale jets},} in IAU Symposium, Vol. 313, Extragalactic Jets from Every Angle, ed. F.~{Massaro}, C.~C. {Cheung}, E.~{Lopez}, \& A.~{Siemiginowska}, 108--115, \dodoi{10.1017/S1743921315002021}

\bibitem[{A. {Lalakos} {et~al.}(2022){Lalakos}, {Gottlieb}, {Kaaz}, {Chatterjee}, {Liska}, {Christie}, {Tchekhovskoy}, {Zhuravleva}, \& {Nokhrina}}]{Lalakos2022}
{Lalakos}, A., {Gottlieb}, O., {Kaaz}, N., {et~al.} 2022, \bibinfo{title}{{Bridging the Bondi and Event Horizon Scales: 3D GRMHD Simulations Reveal X-shaped Radio Galaxy Morphology},} \apjl, 936, L5, \dodoi{10.3847/2041-8213/ac7bed}

\bibitem[{L. {Lovisari} {et~al.}(2020){Lovisari}, {Ettori}, {Sereno}, {Schellenberger}, {Forman}, {Andrade-Santos}, \& {Jones}}]{Lovisari2020}
{Lovisari}, L., {Ettori}, S., {Sereno}, M., {et~al.} 2020, \bibinfo{title}{{Comparing different mass estimators for a large subsample of the Planck-ESZ clusters},} \aap, 644, A78, \dodoi{10.1051/0004-6361/202038718}

\bibitem[{C.-J. {Ma} {et~al.}(2013){Ma}, {McNamara}, \& {Nulsen}}]{Ma2013}
{Ma}, C.-J., {McNamara}, B.~R., \& {Nulsen}, P.~E.~J. 2013, \bibinfo{title}{{Radio Active Galaxy Nuclei in Galaxy Clusters: Heating Hot Atmospheres and Driving Supermassive Black Hole Growth over Cosmic Time},} \apj, 763, 63, \dodoi{10.1088/0004-637X/763/1/63}

\bibitem[{J. {Machalski} {et~al.}(2008){Machalski}, {Kozie{\l}-Wierzbowska}, {Jamrozy}, \& {Saikia}}]{Machalski2008}
{Machalski}, J., {Kozie{\l}-Wierzbowska}, D., {Jamrozy}, M., \& {Saikia}, D.~J. 2008, \bibinfo{title}{{J1420-0545: The Radio Galaxy Larger than 3C 236},} \apj, 679, 149, \dodoi{10.1086/586703}

\bibitem[{M. {Mahato} {et~al.}(2026){Mahato}, {Tempel}, {Sankhyayan}, {Dabhade}, \& {Chavan}}]{Mahato2026}
{Mahato}, M., {Tempel}, E., {Sankhyayan}, S., {Dabhade}, P., \& {Chavan}, K. 2026, \bibinfo{title}{{Search and analysis of giant radio galaxies with associated nuclei (SAGAN): VI: When jets meet filaments - Environmental imprints on the growth of giant radio galaxies},} \aap, 706, A310, \dodoi{10.1051/0004-6361/202557646}

\bibitem[{J.~M. {Malarecki} {et~al.}(2015){Malarecki}, {Jones}, {Saripalli}, {Staveley-Smith}, \& {Subrahmanyan}}]{Malarecki2015}
{Malarecki}, J.~M., {Jones}, D.~H., {Saripalli}, L., {Staveley-Smith}, L., \& {Subrahmanyan}, R. 2015, \bibinfo{title}{{Giant radio galaxies - II. Tracers of large-scale structure},} \mnras, 449, 955, \dodoi{10.1093/mnras/stv273}

\bibitem[{M. {Meenakshi} {et~al.}(2023){Meenakshi}, {Mukherjee}, {Bodo}, \& {Rossi}}]{Meenakshi2023}
{Meenakshi}, M., {Mukherjee}, D., {Bodo}, G., \& {Rossi}, P. 2023, \bibinfo{title}{{A polarization study of jets interacting with turbulent magnetic fields},} \mnras, 526, 5418, \dodoi{10.1093/mnras/stad3092}

\bibitem[{M. {Meenakshi} {et~al.}(2024){Meenakshi}, {Mukherjee}, {Bodo}, {Rossi}, \& {Harrison}}]{Meenakshi2024}
{Meenakshi}, M., {Mukherjee}, D., {Bodo}, G., {Rossi}, P., \& {Harrison}, C.~M. 2024, \bibinfo{title}{{A comparative study of radio signatures from winds and jets: modelling synchrotron emission and polarization},} \mnras, 533, 2213, \dodoi{10.1093/mnras/stae1890}

\bibitem[{A. {Mignone} {et~al.}(2007){Mignone}, {Bodo}, {Massaglia}, {Matsakos}, {Tesileanu}, {Zanni}, \& {Ferrari}}]{Mignone2007}
{Mignone}, A., {Bodo}, G., {Massaglia}, S., {et~al.} 2007, \bibinfo{title}{{PLUTO: A Numerical Code for Computational Astrophysics},} \apjs, 170, 228, \dodoi{10.1086/513316}

\bibitem[{A. {Mignone} \& J.~C. {McKinney}(2007){Mignone} \& {McKinney}}]{Mignone2007_EoS}
{Mignone}, A., \& {McKinney}, J.~C. 2007, \bibinfo{title}{{Equation of state in relativistic magnetohydrodynamics: variable versus constant adiabatic index},} \mnras, 378, 1118, \dodoi{10.1111/j.1365-2966.2007.11849.x}

\bibitem[{Y. {Mizuno} {et~al.}(2015){Mizuno}, {G{\'o}mez}, {Nishikawa}, {Meli}, {Hardee}, \& {Rezzolla}}]{Mizuno2015}
{Mizuno}, Y., {G{\'o}mez}, J.~L., {Nishikawa}, K.-I., {et~al.} 2015, \bibinfo{title}{{Recollimation Shocks in Magnetized Relativistic Jets},} \apj, 809, 38, \dodoi{10.1088/0004-637X/809/1/38}

\bibitem[{R.~I.~J. {Mostert} {et~al.}(2024){Mostert}, {Oei}, {Barkus}, {Alegre}, {Hardcastle}, {Duncan}, {R{\"o}ttgering}, {van Weeren}, \& {Horton}}]{Mostert2024}
{Mostert}, R.~I.~J., {Oei}, M.~S.~S.~L., {Barkus}, B., {et~al.} 2024, \bibinfo{title}{{Constraining the giant radio galaxy population with machine learning and Bayesian inference},} \aap, 691, A185, \dodoi{10.1051/0004-6361/202348897}

\bibitem[{D. {Mukherjee} {et~al.}(2016){Mukherjee}, {Bicknell}, {Sutherland}, \& {Wagner}}]{Mukherjee2016}
{Mukherjee}, D., {Bicknell}, G.~V., {Sutherland}, R., \& {Wagner}, A. 2016, \bibinfo{title}{{Relativistic jet feedback in high-redshift galaxies - I. Dynamics},} \mnras, 461, 967, \dodoi{10.1093/mnras/stw1368}

\bibitem[{D. {Mukherjee} {et~al.}(2020){Mukherjee}, {Bodo}, {Mignone}, {Rossi}, \& {Vaidya}}]{Mukherjee2020}
{Mukherjee}, D., {Bodo}, G., {Mignone}, A., {Rossi}, P., \& {Vaidya}, B. 2020, \bibinfo{title}{{Simulating the dynamics and non-thermal emission of relativistic magnetized jets I. Dynamics},} \mnras, 499, 681, \dodoi{10.1093/mnras/staa2934}

\bibitem[{A. {Neronov} {et~al.}(2025){Neronov}, {Vazza}, {Brandenburg}, \& {Caprini}}]{Neronov2025}
{Neronov}, A., {Vazza}, F., {Brandenburg}, A., \& {Caprini}, C. 2025, \bibinfo{title}{{Intergalactic magnetism in a {\ensuremath{\gamma}}-ray beam as a model of Porphyrion},} \aap, 696, L8, \dodoi{10.1051/0004-6361/202453099}

\bibitem[{E.~F. {Ocran} {et~al.}(2021){Ocran}, {Taylor}, {Vaccari}, {Ishwara-Chandra}, {Prandoni}, {Prescott}, \& {Mancuso}}]{Ocran2021}
{Ocran}, E.~F., {Taylor}, A.~R., {Vaccari}, M., {et~al.} 2021, \bibinfo{title}{{The evolution of the low-frequency radio AGN population to z {$\simeq$} 1.5 in the ELAIS N1 field},} \mnras, 500, 4685, \dodoi{10.1093/mnras/staa3538}

\bibitem[{M.~S.~S.~L. {Oei} {et~al.}(2022){Oei}, {van Weeren}, {Hardcastle}, {Botteon}, {Shimwell}, {Dabhade}, {Gast}, {R{\"o}ttgering}, {Br{\"u}ggen}, {Tasse}, {Williams}, \& {Shulevski}}]{Oei2022_5Mpc}
{Oei}, M. S.~S.~L., {van Weeren}, R.~J., {Hardcastle}, M.~J., {et~al.} 2022, \bibinfo{title}{{The discovery of a radio galaxy of at least 5 Mpc},} \aap, 660, A2, \dodoi{10.1051/0004-6361/202142778}

\bibitem[{M.~S.~S.~L. {Oei} {et~al.}(2023){Oei}, {van Weeren}, {Gast}, {Botteon}, {Hardcastle}, {Dabhade}, {Shimwell}, {R{\"o}ttgering}, \& {Drabent}}]{Oei2023}
{Oei}, M. S.~S.~L., {van Weeren}, R.~J., {Gast}, A. R.~D.~J.~G.~I.~B., {et~al.} 2023, \bibinfo{title}{{Measuring the giant radio galaxy length distribution with the LoTSS},} \aap, 672, A163, \dodoi{10.1051/0004-6361/202243572}

\bibitem[{M.~S.~S.~L. {Oei} {et~al.}(2024){Oei}, {Hardcastle}, {Timmerman}, {Gast}, {Botteon}, {Rodriguez}, {Stern}, {Calistro Rivera}, {van Weeren}, {R{\"o}ttgering}, {Intema}, {de Gasperin}, \& {Djorgovski}}]{Oei2024_7Mpc}
{Oei}, M. S.~S.~L., {Hardcastle}, M.~J., {Timmerman}, R., {et~al.} 2024, \bibinfo{title}{{Black hole jets on the scale of the cosmic web},} \nat, 633, 537, \dodoi{10.1038/s41586-024-07879-y}

\bibitem[{S. {Paul} {et~al.}(2017){Paul}, {John}, {Gupta}, \& {Kumar}}]{Paul2017}
{Paul}, S., {John}, R.~S., {Gupta}, P., \& {Kumar}, H. 2017, \bibinfo{title}{{Understanding `galaxy groups' as a unique structure in the universe},} \mnras, 471, 2, \dodoi{10.1093/mnras/stx1488}

\bibitem[{M. {Perucho} {et~al.}(2019){Perucho}, {Mart{\'\i}}, \& {Quilis}}]{Perucho2019}
{Perucho}, M., {Mart{\'\i}}, J.-M., \& {Quilis}, V. 2019, \bibinfo{title}{{Long-term FRII jet evolution: clues from three-dimensional simulations},} \mnras, 482, 3718, \dodoi{10.1093/mnras/sty2912}

\bibitem[{P. {Ratnaparkhi} {et~al.}(2026){Ratnaparkhi}, {Mukherjee}, {Bodo}, {Rossi}, {Brienza}, \& {Morganti}}]{Ratnaparkhi2026}
{Ratnaparkhi}, P., {Mukherjee}, D., {Bodo}, G., {et~al.} 2026, \bibinfo{title}{{Numerical Simulations of Restarted Jets -- I. Dynamics},} arXiv e-prints, arXiv:2607.03061.
\newblock \doarXiv{2607.03061}

\bibitem[{M. Rossazza {et~al.}(2026)Rossazza, Mignone, Bugli, Truzzi, Riha, Panoc, Vysocky, Shukla, Romeo, \& Berta}]{Rossazza2025}
Rossazza, M., Mignone, A., Bugli, M., {et~al.} 2026, \bibinfo{title}{The PLUTO code on GPUs: A first look at Eulerian MHD methods,} Astronomy and Computing, 55, 101076, \dodoi{https://doi.org/10.1016/j.ascom.2026.101076}

\bibitem[{P. {Rossi} {et~al.}(2024){Rossi}, {Bodo}, {Massaglia}, \& {Capetti}}]{Rossi2024}
{Rossi}, P., {Bodo}, G., {Massaglia}, S., \& {Capetti}, A. 2024, \bibinfo{title}{{The different flavors of extragalactic jets: Magnetized relativistic flows},} \aap, 685, A4, \dodoi{10.1051/0004-6361/202348864}

\bibitem[{V. {Safouris} {et~al.}(2009){Safouris}, {Subrahmanyan}, {Bicknell}, \& {Saripalli}}]{Safouris2009}
{Safouris}, V., {Subrahmanyan}, R., {Bicknell}, G.~V., \& {Saripalli}, L. 2009, \bibinfo{title}{{MRCB0319-454: probing the large-scale structure with a giant radio galaxy},} \mnras, 393, 2, \dodoi{10.1111/j.1365-2966.2008.14181.x}

\bibitem[{S. {Sankhyayan} \& P. {Dabhade}(2024){Sankhyayan} \& {Dabhade}}]{Sankhyayan2024}
{Sankhyayan}, S., \& {Dabhade}, P. 2024, \bibinfo{title}{{Search and analysis of giant radio galaxies with associated nuclei (SAGAN). IV. Interplay with the Supercluster environment},} \aap, 687, L8, \dodoi{10.1051/0004-6361/202450011}

\bibitem[{L. {Saripalli} {et~al.}(1994){Saripalli}, {Subrahmanyan}, \& {Hunstead}}]{Saripalli1994}
{Saripalli}, L., {Subrahmanyan}, R., \& {Hunstead}, R.~W. 1994, \bibinfo{title}{{0319-454 : an FR II giant radio galaxy with twin jets.},} \mnras, 269, 37, \dodoi{10.1093/mnras/269.1.37}

\bibitem[{A.~P. {Schoenmakers} {et~al.}(2000){Schoenmakers}, {Mack}, {de Bruyn}, {R{\"o}ttgering}, {Klein}, \& {van der Laan}}]{Schoenmakers2000}
{Schoenmakers}, A.~P., {Mack}, K.-H., {de Bruyn}, A.~G., {et~al.} 2000, \bibinfo{title}{{A new sample of giant radio galaxies from the WENSS survey. II. A multi-frequency radio study of a complete sample: Properties of the radio lobes and their environment},} \aaps, 146, 293, \dodoi{10.1051/aas:2000267}

\bibitem[{B. {Sebastian} {et~al.}(2018){Sebastian}, {Ishwara-Chandra}, {Joshi}, \& {Wadadekar}}]{Sebastian2018}
{Sebastian}, B., {Ishwara-Chandra}, C.~H., {Joshi}, R., \& {Wadadekar}, Y. 2018, \bibinfo{title}{{Discovery of a new, 2.2-Mpc giant radio galaxy at a redshift of 0.57},} \mnras, 473, 4926, \dodoi{10.1093/mnras/stx2631}

\bibitem[{T.~W. {Shimwell} {et~al.}(2026){Shimwell}, {Hardcastle}, {Tasse}, {Drabent}, {Botteon}, {Williams}, {Best}, {R{\"o}ttgering}, {Br{\"u}ggen}, {Brunetti}, {Callingham}, {Chy{\.z}y}, {Conway}, {De Gasperin}, {Haverkorn}, {Horellou}, {Jackson}, {Miley}, {Morabito}, {Morganti}, {O'Sullivan}, {Schwarz}, {Smith}, {van Weeren}, {Vedantham}, {White}, {Ahmadi}, {Alegre}, {Arias}, {Asabere}, {Bahr-Kalus}, {Barkus}, {Bilicki}, {B{\"o}hme}, {Brentjens}, {Brienza}, {Bomans}, {Bonafede}, {Bonato}, {Bonnassieux}, {Boxelaar}, {Camera}, {Cassano}, {Chilufya}, {Cianfaglione}, {Croston}, {Cuciti}, {Dabhade}, {De Rubeis}, {de Jong}, {Dallacasa}, {Dettmar}, {Duncan}, {Di Gennaro}, {Edler}, {Groeneveld}, {G{\"u}rkan}, {Hajduk}, {Hale}, {Heesen}, {Hoang}, {Hoeft}, {Holties}, {Horton}, {Iacobelli}, {Jamrozy}, {Jarvis}, {Jelic}, {Kadler}, {Kondapally}, {Kunert-Bajraszewska}, {Loose}, {Magliocchetti}, {Ma{\l}ek}, {Manzano}, {McKean}, {Mevius}, {Mingo}, {Miskolczi}, {Misra}, {Mold{\'o}n}, {Nair}, {Nakoneczny}, {Orru},
  {Pashapour-Ahmadabadi}, {Pasini}, {Petley}, {Pierce}, {Prandoni}, {Rafferty}, {Rajpurohit}, {Riseley}, {Roberts}, {Sethi}, {Shulevski}, {Stein}, {Stuardi}, {Sweijen}, {ter Veen}, {Timmerman}, {Vaccari}, \& {Wijnholds}}]{Shimwell2026}
{Shimwell}, T.~W., {Hardcastle}, M.~J., {Tasse}, C., {et~al.} 2026, \bibinfo{title}{{The LOFAR Two-metre Sky Survey: VII. Third Data Release},} \aap, 707, A198, \dodoi{10.1051/0004-6361/202557749}

\bibitem[{M. {Simonte} {et~al.}(2024){Simonte}, {Andernach}, {Br{\"u}ggen}, {Miley}, \& {Barthel}}]{Simonte2024}
{Simonte}, M., {Andernach}, H., {Br{\"u}ggen}, M., {Miley}, G.~K., \& {Barthel}, P. 2024, \bibinfo{title}{{Giant radio galaxies in the LOFAR deep fields},} \aap, 686, A21, \dodoi{10.1051/0004-6361/202348904}

\bibitem[{M. {Sun} {et~al.}(2009){Sun}, {Voit}, {Donahue}, {Jones}, {Forman}, \& {Vikhlinin}}]{Sun2009}
{Sun}, M., {Voit}, G.~M., {Donahue}, M., {et~al.} 2009, \bibinfo{title}{{Chandra Studies of the X-Ray Gas Properties of Galaxy Groups},} \apj, 693, 1142, \dodoi{10.1088/0004-637X/693/2/1142}

\bibitem[{H. {Tang} {et~al.}(2020){Tang}, {Scaife}, {Wong}, {Kapi{\'n}ska}, {Rudnick}, {Shabala}, {Seymour}, \& {Norris}}]{Tang2020}
{Tang}, H., {Scaife}, A.~M.~M., {Wong}, O.~I., {et~al.} 2020, \bibinfo{title}{{Radio Galaxy Zoo: new giant radio galaxies in the RGZ DR1 catalogue},} \mnras, 499, 68, \dodoi{10.1093/mnras/staa2805}

\bibitem[{A. {Tchekhovskoy} \& O. {Bromberg}(2016){Tchekhovskoy} \& {Bromberg}}]{Tchekhovskoy2016}
{Tchekhovskoy}, A., \& {Bromberg}, O. 2016, \bibinfo{title}{{Three-dimensional relativistic MHD simulations of active galactic nuclei jets: magnetic kink instability and Fanaroff-Riley dichotomy},} \mnras, 461, L46, \dodoi{10.1093/mnrasl/slw064}

\bibitem[{M. {Turisini} {et~al.}(2023){Turisini}, {Amati}, \& {Cestari}}]{Turisini2023}
{Turisini}, M., {Amati}, G., \& {Cestari}, M. 2023, \bibinfo{title}{{LEONARDO: A Pan-European Pre-Exascale Supercomputer for HPC and AI Applications},} arXiv e-prints, arXiv:2307.16885, \dodoi{10.48550/arXiv.2307.16885}

\bibitem[{N. {Upreti} {et~al.}(2024){Upreti}, {Vaidya}, \& {Shukla}}]{Upreti2024}
{Upreti}, N., {Vaidya}, B., \& {Shukla}, A. 2024, \bibinfo{title}{{Bridging simulations of kink instability in relativistic magnetized jets with radio emission and polarisation},} Journal of High Energy Astrophysics, 44, 146, \dodoi{10.1016/j.jheap.2024.09.007}

\bibitem[{B. {Vaidya} {et~al.}(2018){Vaidya}, {Mignone}, {Bodo}, {Rossi}, \& {Massaglia}}]{Vaidya2018}
{Vaidya}, B., {Mignone}, A., {Bodo}, G., {Rossi}, P., \& {Massaglia}, S. 2018, \bibinfo{title}{{A Particle Module for the PLUTO Code. II. Hybrid Framework for Modeling Nonthermal Emission from Relativistic Magnetized Flows},} \apj, 865, 144, \dodoi{10.3847/1538-4357/aadd17}

\bibitem[{I.~P. {van der Westhuizen} {et~al.}(2019){van der Westhuizen}, {van Soelen}, {Meintjes}, \& {Beall}}]{VanDer2019}
{van der Westhuizen}, I.~P., {van Soelen}, B., {Meintjes}, P.~J., \& {Beall}, J.~H. 2019, \bibinfo{title}{{Using synchrotron emission modelling of relativistic hydrodynamic jet simulations to study the FR I/FR II dichotomy of active galactic nuclei radio jets},} \mnras, 485, 4658, \dodoi{10.1093/mnras/stz734}

\bibitem[{M.~P. {van Haarlem} {et~al.}(2013){van Haarlem}, {Wise}, {Gunst}, {Heald}, {McKean}, {Hessels}, {de Bruyn}, {Nijboer}, {Swinbank}, {Fallows}, {Brentjens}, {Nelles}, {Beck}, {Falcke}, {Fender}, {H{\"o}randel}, {Koopmans}, {Mann}, {Miley}, {R{\"o}ttgering}, {Stappers}, {Wijers}, {Zaroubi}, {van den Akker}, {Alexov}, {Anderson}, {Anderson}, {van Ardenne}, {Arts}, {Asgekar}, {Avruch}, {Batejat}, {B{\"a}hren}, {Bell}, {Bell}, {van Bemmel}, {Bennema}, {Bentum}, {Bernardi}, {Best}, {B{\^\i}rzan}, {Bonafede}, {Boonstra}, {Braun}, {Bregman}, {Breitling}, {van de Brink}, {Broderick}, {Broekema}, {Brouw}, {Br{\"u}ggen}, {Butcher}, {van Cappellen}, {Ciardi}, {Coenen}, {Conway}, {Coolen}, {Corstanje}, {Damstra}, {Davies}, {Deller}, {Dettmar}, {van Diepen}, {Dijkstra}, {Donker}, {Doorduin}, {Dromer}, {Drost}, {van Duin}, {Eisl{\"o}ffel}, {van Enst}, {Ferrari}, {Frieswijk}, {Gankema}, {Garrett}, {de Gasperin}, {Gerbers}, {de Geus}, {Grie{\ss}meier}, {Grit}, {Gruppen}, {Hamaker}, {Hassall}, {Hoeft}, {Holties},
  {Horneffer}, {van der Horst}, {van Houwelingen}, {Huijgen}, {Iacobelli}, {Intema}, {Jackson}, {Jelic}, {de Jong}, {Juette}, {Kant}, {Karastergiou}, {Koers}, {Kollen}, {Kondratiev}, {Kooistra}, {Koopman}, {Koster}, {Kuniyoshi}, {Kramer}, {Kuper}, {Lambropoulos}, {Law}, {van Leeuwen}, {Lemaitre}, {Loose}, {Maat}, {Macario}, {Markoff}, {Masters}, {McFadden}, {McKay-Bukowski}, {Meijering}, {Meulman}, {Mevius}, {Middelberg}, {Millenaar}, {Miller-Jones}, {Mohan}, {Mol}, {Morawietz}, {Morganti}, {Mulcahy}, {Mulder}, {Munk}, {Nieuwenhuis}, {van Nieuwpoort}, {Noordam}, {Norden}, {Noutsos}, {Offringa}, {Olofsson}, {Omar}, {Orr{\'u}}, {Overeem}, {Paas}, {Pandey-Pommier}, {Pandey}, {Pizzo}, {Polatidis}, {Rafferty}, {Rawlings}, {Reich}, {de Reijer}, {Reitsma}, {Renting}, {Riemers}, {Rol}, {Romein}, {Roosjen}, {Ruiter}, {Scaife}, {van der Schaaf}, {Scheers}, {Schellart}, {Schoenmakers}, {Schoonderbeek}, {Serylak}, {Shulevski}, {Sluman}, {Smirnov}, {Sobey}, {Spreeuw}, {Steinmetz}, {Sterks}, {Stiepel}, {Stuurwold},
  {Tagger}, {Tang}, {Tasse}, {Thomas}, {Thoudam}, {Toribio}, {van der Tol}, {Usov}, {van Veelen}, {van der Veen}, {ter Veen}, {Verbiest}, {Vermeulen}, {Vermaas}, {Vocks}, {Vogt}, {de Vos}, {van der Wal}, {van Weeren}, {Weggemans}, {Weltevrede}, {White}, {Wijnholds}, {Wilhelmsson}, {Wucknitz}, {Yatawatta}, {Zarka}, \& {Zensus}}]{vanHaarlem2013}
{van Haarlem}, M.~P., {Wise}, M.~W., {Gunst}, A.~W., {et~al.} 2013, \bibinfo{title}{{LOFAR: The LOw-Frequency ARray},} \aap, 556, A2, \dodoi{10.1051/0004-6361/201220873}

\bibitem[{F. {Vazza} \& G. {Brunetti}(2026){Vazza} \& {Brunetti}}]{Vazza2026}
{Vazza}, F., \& {Brunetti}, G. 2026, \bibinfo{title}{{The interpretation of XRISM X-ray measurements of turbulence in the intracluster medium: A comparison with cosmological simulations},} \aap, 705, A129, \dodoi{10.1051/0004-6361/202556070}

\bibitem[{S. {Walg} {et~al.}(2020){Walg}, {Achterberg}, {Markoff}, {Keppens}, \& {Porth}}]{Walg2020}
{Walg}, S., {Achterberg}, A., {Markoff}, S., {Keppens}, R., \& {Porth}, O. 2020, \bibinfo{title}{{Relativistic AGN jets - III. Synthesis of synchrotron emission from double-double radio galaxies},} \mnras, 497, 3638, \dodoi{10.1093/mnras/staa2195}

\bibitem[{J.-S. {Wang} {et~al.}(2023){Wang}, {Reville}, {Mizuno}, {Rieger}, \& {Aharonian}}]{Wang2023}
{Wang}, J.-S., {Reville}, B., {Mizuno}, Y., {Rieger}, F.~M., \& {Aharonian}, F.~A. 2023, \bibinfo{title}{{Particle acceleration in shearing flows: the self-generation of turbulent spine-sheath structures in relativistic magnetohydrodynamic jet simulations},} \mnras, 519, 1872, \dodoi{10.1093/mnras/stac3616}

\bibitem[{A.~G. {Willis} {et~al.}(1974){Willis}, {Strom}, \& {Wilson}}]{Willis1974}
{Willis}, A.~G., {Strom}, R.~G., \& {Wilson}, A.~S. 1974, \bibinfo{title}{{3C236, DA240; the largest radio sources known},} \nat, 250, 625, \dodoi{10.1038/250625a0}

\bibitem[{ {Xrism Collaboration} {et~al.}(2025){Xrism Collaboration}, {Audard}, {Awaki}, {Ballhausen}, {Bamba}, {Behar}, {Boissay-Malaquin}, {Brenneman}, {Brown}, {Corrales}, {Costantini}, {Cumbee}, {Diaz Trigo}, {Done}, {Dotani}, {Ebisawa}, {Eckart}, {Eckert}, {Eguchi}, {Enoto}, {Ezoe}, {Foster}, {Fujimoto}, {Fujita}, {Fukazawa}, {Fukushima}, {Furuzawa}, {Gallo}, {Garc{\'\i}a}, {Gu}, {Guainazzi}, {Hagino}, {Hamaguchi}, {Hatsukade}, {Hayashi}, {Hayashi}, {Hell}, {Hodges-Kluck}, {Hornschemeier}, {Ichinohe}, {Ishi}, {Ishida}, {Ishikawa}, {Ishisaki}, {Kaastra}, {Kallman}, {Kara}, {Katsuda}, {Kanemaru}, {Kelley}, {Kilbourne}, {Kitamoto}, {Kobayashi}, {Kohmura}, {Kubota}, {Leutenegger}, {Loewenstein}, {Maeda}, {Markevitch}, {Matsumoto}, {Matsushita}, {McCammon}, {McNamara}, {Mernier}, {Miller}, {Miller}, {Mitsuishi}, {Mizumoto}, {Mizuno}, {Mori}, {Mukai}, {Murakami}, {Mushotzky}, {Nakajima}, {Nakazawa}, {Ness}, {Nobukawa}, {Nobukawa}, {Noda}, {Odaka}, {Ogawa}, {Ogorza{\l}ek}, {Okajima}, {Ota}, {Paltani}, {Petre},
  {Plucinsky}, {Porter}, {Pottschmidt}, {Sato}, {Sato}, {Sawada}, {Seta}, {Shidatsu}, {Simionescu}, {Smith}, {Suzuki}, {Szymkowiak}, {Takahashi}, {Takeo}, {Tamagawa}, {Tamura}, {Tanaka}, {Tanimoto}, {Tashiro}, {Terada}, {Terashima}, {Tsuboi}, {Tsujimoto}, {Tsunemi}, {Tsuru}, {T{\"u}mer}, {Uchida}, {Uchida}, {Uchida}, {Uchiyama}, {Ueda}, {Ueda}, {Uno}, {Vink}, {Watanabe}, {Williams}, {Yamada}, {Yamada}, {Yamaguchi}, {Yamaoka}, {Yamasaki}, {Yamauchi}, {Yamauchi}, {Yaqoob}, {Yoneyama}, {Yoshida}, {Yukita}, {Zhuravleva}, {Fabian}, {Nelson}, {Okabe}, {Pillepich}, {Potter}, {Regamey}, {Sakai}, {Shishido}, {Truong}, {Wik}, \& {Zuhone}}]{Xrism2025}
{Xrism Collaboration}, {Audard}, M., {Awaki}, H., {et~al.} 2025, \bibinfo{title}{{XRISM Forecast for the Coma Cluster: Stormy, with a Steep Power Spectrum},} \apjl, 985, L20, \dodoi{10.3847/2041-8213/add2f6}

\bibitem[{S.~A. {Young} {et~al.}(2025){Young}, {Turner}, {Shabala}, {Stewart}, \& {Yates-Jones}}]{Young2025}
{Young}, S.~A., {Turner}, R.~J., {Shabala}, S.~S., {Stewart}, G. S.~C., \& {Yates-Jones}, P.~M. 2025, \bibinfo{title}{{Spectral signatures of young radio galaxies},} \pasa, 42, e100, \dodoi{10.1017/pasa.2025.10068}

\end{thebibliography}
\bibliographystyle{aasjournalv7}



\end{document}